\documentclass[8.5pt,twoside,twocolumn]{article}
\usepackage[super,sort&compress,comma]{natbib} 
\usepackage{mhchem}
\usepackage{times,mathptmx}
\usepackage{sectsty}
\usepackage{balance} 

\usepackage{graphicx} %eps figures can be used instead
\usepackage{lastpage}
\usepackage[format=plain,justification=raggedright,singlelinecheck=false,font=small,labelfont=bf,labelsep=space]{caption} 
\usepackage{fancyhdr}
\usepackage{amssymb}
\usepackage{color}
\usepackage{wasysym}
\definecolor{green0}{rgb}{0,0.5,0}

\begin{document}

\thispagestyle{plain}
\fancypagestyle{plain}{
%\fancyhead[L]{\includegraphics[height=8pt]{headers/LH}}
%\fancyhead[C]{\hspace{-1cm}\includegraphics[height=20pt]{headers/CH}}
%\fancyhead[R]{\includegraphics[height=10pt]{headers/RH}\vspace{-0.2cm}}
\renewcommand{\headrulewidth}{1pt}}
\renewcommand{\thefootnote}{\fnsymbol{footnote}}
\renewcommand\footnoterule{\vspace*{1pt}% 
\hrule width 3.4in height 0.4pt \vspace*{5pt}} 
\setcounter{secnumdepth}{5}

\makeatletter 
\def\subsubsection{\@startsection{subsubsection}{3}{10pt}{-1.25ex plus -1ex minus -.1ex}{0ex plus 0ex}{\normalsize\bf}} 
\def\paragraph{\@startsection{paragraph}{4}{10pt}{-1.25ex plus -1ex minus -.1ex}{0ex plus 0ex}{\normalsize\textit}} 
\renewcommand\@biblabel[1]{#1}            
\renewcommand\@makefntext[1]% 
{\noindent\makebox[0pt][r]{\@thefnmark\,}#1}
\makeatother 
\renewcommand{\figurename}{\small{Fig.}~}
\sectionfont{\large}
\subsectionfont{\normalsize} 

\fancyfoot{}
%\fancyfoot[LO,RE]{\vspace{-7pt}\includegraphics[height=9pt]{headers/LF}}
%\fancyfoot[CO]{\vspace{-7.2pt}\hspace{12.2cm}\includegraphics{headers/RF}}
%\fancyfoot[CE]{\vspace{-7.5pt}\hspace{-13.5cm}\includegraphics{headers/RF}}
%\fancyfoot[RO]{\footnotesize{\sffamily{1--\pageref{LastPage} ~\textbar  \hspace{2pt}\thepage}}}
%\fancyfoot[LE]{\footnotesize{\sffamily{\thepage~\textbar\hspace{3.45cm} 1--\pageref{LastPage}}}}
\fancyhead{}
\renewcommand{\headrulewidth}{1pt} 
\renewcommand{\footrulewidth}{1pt}
\setlength{\arrayrulewidth}{1pt}
\setlength{\columnsep}{6.5mm}
\setlength\bibsep{1pt}

\twocolumn[
  \begin{@twocolumnfalse}
\noindent\LARGE{\textbf{Quantum Entanglement in Carbon-Carbon, Carbon-Phosphorus and Silicon-Silicon Bonds$^\dag$}}
\vspace{0.6cm}

\noindent\large{\textbf{Matthieu Mottet\textit{$^{a}$}, Pawe{\l} Tecmer,$^{\ast}$\textit{$^{a\ddag}$} Katharina Boguslawski,$^{\ast}$\textit{$^{a\ddag}$} \"Ors Legeza,\textit{$^{b}$} and Markus Reiher\textit{$^{a}$}}}\vspace{0.5cm}

\noindent \normalsize{
The chemical bond is an important local concept to understand chemical compounds and processes. 
Unfortunately, like most local concepts, the chemical bond and the bond order do not correspond to any physical observable and thus cannot be determined as an expectation value of a quantum chemical operator. 
We recently demonstrated [Boguslawski \textit{et al., J. Chem. Theory Comput.}, 2013, \textbf{9}, 2959--2973] that one- and two-orbital-based entanglement measures can be applied to interpret electronic wave functions in terms of orbital correlation.
Orbital entanglement emerged to be a powerful tool to provide a qualitative understanding of bond-forming and bond-breaking processes, and allowed for an estimation of bond orders of simple diatomic molecules beyond the classical bonding models. 
In this article we demonstrate that the orbital entanglement analysis can be extended to polyatomic molecules to understand chemical bonding.
}
\vspace{0.5cm}
 \end{@twocolumnfalse}
  ]

% makes reDecayferences listed with 1., 2., etc.   
% \makeatletter
% \renewcommand\@biblabel[1]{#1.}
% \makeatother

%\bibliographystyle{apsrev}
 
%\renewcommand{\baselinestretch}{1.5}
%\normalsize

%\setlength\parindent{0pt}

\section{Introduction} 
\footnotetext{\dag~Submitted to Phys.~Chem.~Chem.~Phys.}
\footnotetext{\textit{$^{a}$~ETH Z\"urich, Laboratory of Physical Chemistry, Vladimir-Prelog-Weg 2, CH-8093 Z\"urich, Switzerland. tecmer@mcmaster.ca, bogusl@mcmaster.ca}}
\footnotetext{\textit{$^{b}$~Strongly Correlated Systems "Lend\"ulet" Research Group, Wigner Research Center for Physics,  H-1525 Budapest, Hungary.}}
%additional addresses can be cited as above using the lower-case letters, c, d, e... If all authors are from the same address, no letter is required
\footnotetext{\ddag~Present address: Department of Chemistry and Chemical Biology, McMaster University, Hamilton, Ontario, Canada.}

Albeit computational chemistry has reached a remarkable level in the quantitative description of
atoms and molecules\cite{bartlett_book_1984,Helgaker_book,bartlett_2007,dft_rev_2012}, these achievements came along with a fading qualitative understanding.
To remedy the receding interpretation of complex total electronic wave functions, quantum chemistry introduced various analysis tools based on local properties, that is properties of atoms or functional groups of atoms in a molecule, like, for instance, bond orders, atomic charges in molecules and local spins~\cite{kutzelnigg_bonding1,Clark2001,
Davidson2001,Clark2002,Herrmann2005,markusjcc2006,Mayer2007,Davidson2007a,markus_fd,Podewitz2008,Alcoba2009,Mayer2009,Podewitz2010,Torre2010,Mayer2010,Alcoba2011,Alcoba2011a,Ramos-Cordoba2012}. Such local concepts allow us to interpret electronic structures in terms of intuitive building blocks and favor a qualitative understanding of chemical processes. 
Similar approaches such as hardness, softness, electronegativity and related properties emerged within conceptual density functional theory~\cite{Geerlings_rev_2003}. 

The bond order~\cite{Wiberg1968,Armstrong1973,Natiello1984,
Mayer1985,Mayer1986,Bochicchio1991a,Alcoba2007,Mayer2007a,Markus-chemistry}
is an important and popular local concept to understand chemical bonding. It constitutes a quantitative measure for the formal number of (covalent and ionic) bonds formed
between two neighboring atoms.
The bond order can elucidate the stability and the strength of a chemical bond and can be thus considered as a guide to chemical synthesis.
Unfortunately, the chemical bond or the bond order cannot be unambiguously assigned to any quantum chemical observable.
As a consequence, various bond order concepts have been introduced in quantum chemistry for both one-determinant~\cite{Mayer1983,Mayer1985,Mayer1986,Mayer1986a} and correlated wave functions~\cite{Alcoba2008,Lobayan2010}.

Conceptually different bond-order measures have been developed for one-determinant wave functions that are based on a simple electron counting scheme~\cite{Herzberg1929,Hall1987,Brynda2006,Roos2007} and thus provide an intuitive and convenient picture of the chemical bond. An extension of the electron-counting scheme to correlated wave functions was presented by Roos \textit{et al.}~\cite{Roos2007} who introduced an effective bond order between two atomic centers that is based on natural occupation numbers of molecular orbitals and thus determined from a one-particle density matrix.
As the natural
occupation numbers may have non-integer values, the resulting effective bond orders will be non-integer, as well.
The true bond order is defined as the lowest integer larger or equal to the
total effective bond order value.

Alternatively, the nature of the chemical bond can be investigated using concepts from quantum information theory~\cite{Nalewajski2000,Nalewajski2004,Nalewajski2009a,Nalewajski2012a,orbitalordering,entanglement_letter,entanglement_bonding_2013}. 
Following the observation that one- and two-orbital-based entanglement measures
provide quantitative means for an assessment and classification of electron
correlation effects among molecular orbitals~\cite{entanglement_letter,CUO_DMRG}, an entanglement analysis on bond-formation and dissociation processes of diatomic molecules~\cite{entanglement_bonding_2013} was presented by some of us. 
Our entanglement study indicated that concepts from quantum information theory can be used to retrieve deeper insights into electronic structures and molecular processes.
Furthermore, entanglement measures based on many-particle reduced density matrices offer a conceptual
understanding of bond-forming and bond-breaking processes. The entanglement
analysis comprises two entropic measures which quantifies the quantum information exchange between molecular orbitals and molecular orbital pairs\cite{,orbitalordering,entanglement_bonding_2013}; (i) The single-orbital
entropy that is determined from the eigenvalues $w_{\alpha,i}$ of the one-orbital reduced
density matrix of a given orbital $i$,

\begin{equation}
	\label{eq:1oentropy}
	s(1)_i=-\sum_{\alpha=1}^4w_{\alpha,i}\ln w_{\alpha,i},
\end{equation}

where $\alpha$ runs over all local states of one molecular orbital (there are 4 in the case of spatial orbitals).
(ii) The mutual information that measures the correlation of two orbitals
embedded in the environment of all other active space orbitals, 

\begin{equation}
	\label{eq:mutualInfo}
	I_{i,j}=\frac{1}{2}[s(1)_i+s(1)_j-s(2)_{i,j}](1-\delta_{ij})
\end{equation}

where $i,j=1\ldots k$ is the orbital index of all $k$ active space orbitals and
$\delta_{ij}$ the Kronecker delta~\cite{Rissler_2006}.
In the above equation, $s(2)_{i,j}$ is the two-orbital entropy and is determined from the eigenvalues
$w_{\alpha,i,j}$ of the two-orbital reduced density matrix $\rho_{i,j}$ of orbital pair $\{i,j\}$ ~\cite{entanglement_bonding_2013},
\begin{equation}
	\label{2oentropy}
	s(2)_{i,j}=-\sum_{\alpha=1}^{16}w_{\alpha,i,j}\ln w_{\alpha,i,j}.
\end{equation}

So far, the entanglement analysis has been applied to the dissociation process of simple 
diatomic molecules, while its application to larger molecular systems has not been investigated, yet.
In this work, we will study the performance of orbital-based entanglement measures in resolving bond-breaking processes and in dissecting electronic structures in terms of entangled molecular orbital building blocks for a number of diatomic and polyatomic molecules.  
Our set of prototypical molecules covers singly, doubly and triply bonded carbon--carbon, carbon--phosphorus and silicon--silicon centers. 
 
\section{Methodology}\label{sec:compdetails}

\subsection{Structure optimization}

All molecular structures were optimized using the Amsterdam Density Functional program package~\cite{adf2012,adf1,adf2} employing the BP86~\cite{Perdew86,Becke} exchange--correlation functional and the Triple-$\zeta$ Polarization basis set~\cite{adf_b}. 
The bond lengths of the two dissociating centers (cf. Table~\ref{tab:SandAS} for equilibrium bond lengths) were systematically varied while the structure of the fragments connected to these centers was relaxed (constrained geometry optimization). 
The optimized structures of all polyatomic molecules can be found in Tables I--VI of the Supporting Information.

\subsection{CASSCF}
\label{sec:casscf}

\begin{table}[h]
\small
\caption{\ Summary of equilibrium bond lengths, $\rm r_e$, Abelian point group symmetries and active space sizes used for CASSCF and DMRG calculations for all molecules. In the standard ($x$,~$y$) notation for active spaces, where $x$ indicates the number of correlated electrons and $y$ number of correlated orbitals.}
\label{tab:SandAS}
		\begin{tabular*}{0.5\textwidth}{@{\extracolsep{\fill}}ccccc}
			\hline
			\hline
			Molecule
			& ${\rm r_e}[{\rm \AA}]$
			& Symmetry
			& CASSCF
			& DMRG
			
		\\ 
			\hline
			C$_2$	
			& ${\rm r_{C-C}}=1.34$		
			& D$_{2h}$
			& ($8,8$)
			& ($8,26$)
		\\
		    $[$CP$]^-$
		    & ${\rm r_{C-P}}=1.62$
			& C$_{2v}$
			& ($10,8$)
			& ($10,26$)
		\\
		    HCP
		    & ${\rm r_{C-P}}=1.55$
			& C$_{2v}$
			& ($10,9$)
			& ($10,26$)
		\\
			C$_2$H$_2$
			& ${\rm r_{C-C}}=1.21$
			& D$_{2h}$
			& ($10,10$)
			& ($10,28$)
		\\
		 
		     C$_2$H$_4$
		    & ${\rm r_{C-C}}=1.34$
			& D$_{2h}$
			& ($12,12$)
			& ($12,28$)
		\\
		   C$_2$H$_6$
		   & ${\rm r_{C-C}}=1.53$
			& C$_{2h}$
			& ($14,14$)
			& ($14,28$)
		\\
			Si$_2$H$_2$
			& ${\rm r_{Si-Si}}=2.11$
			& C$_{2h}$
			& ($10,10$)
			& ($10,26$)
		\\
			Si$_2$H$_4$
			& ${\rm r_{Si-Si}}=2.15$
			& D$_{2h}$
			& ($12,12$)
			& ($12,28$)
		\\
			\hline
			\hline	
		\end{tabular*}
\end{table}
All Complete Active Space Self-Consistent Field~\cite{Roos_casscf,Werner_1985,Knowles_1985} (CASSCF) calculations have been performed using the \textsc{Molpro 2010.1} program package~\cite{molpro}. 
As active space, the full valence shell was taken for all molecular systems (\textit{i.e.} H:
$1s$;
C: $2{s}$, $2{p}_{x}$, $2{p}_{y}$ and $2{p}_{z}$; Si, P: $3{s}$, $3{p}_{x}$, $3{p}_{y}$ and
$3{p}_{z}$). 
For the C$_2$, C$_2$H$_2$, C$_2$H$_4$ and Si$_2$H$_4$ molecules, D$_{2{h}}$ point group
symmetry was imposed, while for
HCP and $[$CP$]^-$ C$_{2{ v}}$ point group symmetry was used. 
For C$_2$H$_6$ and Si$_2$H$_2$, C$_{2{h}}$ point group symmetry was set. 
All CASSCF active space sizes are summarized in Table \ref{tab:SandAS}.

For all molecules but C$_2$, Dunning's cc-pVTZ basis set was used with 
the following contractions H: $(5s,2p,1d)\rightarrow[3s,2p,1d]$; 
C, N: $(10s,5p,2d,1f)\rightarrow[4s,3p,2d,1f]$~\cite{dunning_b}; 
Si, P: $(15s,9p,2d,1f)\rightarrow[5s,4p,2d,1f]$~\cite{dunning_bII}. 
In the case of C$_2$, the large aug-cc-pVTZ basis set (C: $(11s,6p,3d,2f)\rightarrow[5s,4p,3d,2f]$~\cite{dunning_b}) was chosen.

\subsection{DMRG}
\label{sec:dmrg}

All DMRG calculations were performed with the \textsc{Budapest DMRG}
program~\cite{dmrg_ors}. 
As orbital basis, the natural orbitals obtained from the CASSCF calculations as described in the previous 
section were used. 
The active spaces were extended by including additional virtual, Fock-type orbitals for each molecule
as summarized in Table \ref{tab:SandAS}; 
$4 {\rm a}_g$, $2 {\rm b}_{3u}$, $2 {\rm b}_{2u}$, $1 {\rm b}_{1g}$, $4 {\rm b}_{1u}$, $2 {\rm b}_{2g}$, $2 {\rm b}_{3g}$ and $1 {\rm a}_{u}$ for $\rm{C_2}$ and $\rm{C_2H_2}$, $3 {\rm a}_g$, $2 {\rm b}_{3u}$, $2 {\rm b}_{2u}$, $1 {\rm b}_{1g}$, $3 {\rm b}_{1u}$, $2 {\rm b}_{2g}$, $2 {\rm b}_{3g}$ and $1 {\rm a}_{u}$ for Si$_2$H$_4$, 
$3 {\rm a}_g$, $4 {\rm b}_{3u}$, $2 {\rm b}_{2u}$, $2 {\rm b}_{1g}$, $2 {\rm b}_{1u}$, $1 {\rm b}_{2g}$, $1 {\rm b}_{3g}$ and $1 {\rm a}_u$ for ${\rm C_2H_4}$ ,
$8 {\rm a}_1$, $4 {\rm b}_{1}$, $4 {\rm b}_{2}$ and $2 {\rm a}_{2}$ for
CP and HCP (one ${\rm a}_1$ orbital less was added to $\rm{HCP}$), 
$5 {\rm a}_g$, $3 {\rm a}_{u}$, $5 {\rm b}_{u}$ and $3 {\rm b}_{g}$ for $\rm{Si_2H_2}$, 
and finally, $5 {\rm a}_g$, $2 {\rm a}_{u}$, $5 {\rm b}_{u}$ and $2 {\rm b}_{g}$ for $\rm{C_2H_6}$. We refer the reader to Table~\ref{tbl:resolution} for the correspondence of 
irreducible representations of Abelian subgroups chosen in the calculations of linear molecules and the full point group symmetry (D$_{\infty h}$ and C$_{\infty v}$).

To enhance DMRG convergence, the orbital ordering was optimized~\cite{orbitalordering} and
the number of renormalized active-system states was chosen dynamically according to a
predefined threshold value for the quantum information loss~\cite{legeza_dbss}
employing the dynamic block state selection approach~\cite{legeza_dbss3,
legeza_dbss2}. As an initial guess, the dynamically extended active-space (DEAS)
procedure was applied~\cite{legeza_dbss}. 
In the DMRG calculations, the minimum and
maximum number of renormalized active-system states was set to $512$ and $1024$, respectively, while the quantum information
loss $\chi$ was set to $10^{-5}$ in all calculations. 
For the $\rm C_2$ molecule it was necessary to increase the maximal number of renormalized active system states to $2048$.
The notation DMRG($x$,$y$)[$m_{\rm max},m_{\rm min},\chi$] indicates that $x$ electrons are correlated in $y$
orbitals, while the minimal and maximal number of active-system states was set to $m_{\rm min}$ and $m_{\rm max}$, respectively
The minimal number of active-system states during the DEAS initialization procedure was set equal to $m_{\rm max}$ in all calculations.

The convergence behavior of all DMRG calculations with respect to the DMRG parameter set is discussed in the Supporting Information.

\begin{table}[h]
\small
     \caption{Resolution of the relevant irreducible representations of D$_{\infty h}$ and C$_{\infty v}$ point groups in terms of those of the  D$_{2h}$, C$_{2v}$ and C$_{2h}$ subgroups\cite{Altmann}.}
     \label{tbl:resolution}
\begin{tabular*}{0.5\textwidth}{@{\extracolsep{\fill}}cccc|ccccc}
\hline
\hline
D$_{\infty h}$ && D$_{2h}$&& C$_{\infty v}$ &&C$_{2v}$ && C$_{2h}$\\
\hline
$\sigma_g$  &&a$_{g}$                        &&$\sigma$    && a$_{1}$   					&& a$_g$                  \\
$\sigma_u$  &&b$_{1u}$                       &&$\sigma$   && a$_{1}$   					&& b$_u$                  \\
$\pi_g$     && b$_{2g} \oplus$ b$_{3g}$      &&$\pi$  && b$_{1} \oplus$ b$_{2}$       && a$_{g} \oplus$ b$_{g}$ \\
$\pi_u$     && b$_{2u} \oplus $ b$_{3u}$     &&$\pi$    && b$_{1} \oplus$ b$_{2}$       && a$_{u} \oplus$ b$_{u}$ \\
$\delta_g$  && a$_{g} \oplus$ b$_{1g}$       &&$\delta$   && a$_{1} \oplus$ a$_{2}$       && a$_{g} \oplus$ b$_{g}$ \\
$\delta_u$  && a$_{u} \oplus$ b$_{1u}$       &&$\delta$    && a$_{1} \oplus$ a$_{2}$       && a$_{u} \oplus$ b$_{u}$\\
$\phi_g$    &&b$_{2g} \oplus$ b$_{3g}$       &&$\phi$    && b$_{1} \oplus$ b$_{2}$       && a$_{g} \oplus$ b$_{g}$ \\
$\phi_u$    &&b$_{2u} \oplus $ b$_{3u}$      &&$\phi$    && b$_{1} \oplus$ b$_{2}$       && a$_{u} \oplus$ b$_{u}$   \\
\hline
\hline
\end{tabular*}
\end{table}

\section{Interpretation of entanglement diagrams}\label{sec:interpret}

To illustrate the entanglement measures, we use a diagrammatic representation of the mutual information and single-orbital entropies as discussed in Refs.~\citenum{entanglement_letter,entanglement_bonding_2013}. The mutual information $I_{i,j}$ between each orbital pair $\{i,j\}$ is color-coded; strongly entangled orbitals are connected via blue lines ($I_{i,j}\approx 10^{-1}$), moderately entangled orbitals are linked by red lines ($I_{i,j}\approx 10^{-2}$), while weakly entangled orbital pairs are indicated by green lines ($I_{i,j}\approx 10^{-3}$). 

The single-orbital entropies, Eq.~\eqref{eq:1oentropy}, and mutual information, Eq.~\eqref{eq:mutualInfo}, corresponding to molecular orbitals associated with a chemical bond, \emph{i.e.}, the bonding and antibonding orbital pairs, are large ($s(1)_i > 0.4$ and $I_{i,j} \approx 10^{-1}$) when bonds are stretched, while the remaining (active space) orbitals are only moderately ($0.1 < s(1)_i < 0.4$ and $I_{i,j} \approx 10^{-2}$) or weakly entangled ($s(1)_i < 0.1$ and $I_{i,j} \approx 10^{-3}$)~\cite{entanglement_bonding_2013}. 
Thus, in the dissociation limit, each bonding and antibonding orbital pair that is important to describe chemical bonding between two centers becomes strongly entangled and renders the dominant contributions in the $I_{i,j}$ and $s(1)_i$ diagrams. 
This observation provides a qualitative picture of the bond multiplicity between two centers. An entanglement-based (or entropic) bond order can be determined from the total number of steep changes in the $s(1)_i$-diagram observed in the dissociation limit after dividing by two to account for the bonding and antibonding combinations of molecular orbitals. 

Furthermore, the rate of growth in $s(1)_i$ depends on the type (or entanglement strength) of a specific bond as the one- and two-orbital entanglement measures are derived from the electronic wave function~\cite{entanglement_bonding_2013}. For instance, ${s(1)_i}$ of orbitals involved in weak $\pi$-bonds increase faster than those corresponding to strong $\sigma$-bonds. A chemical bond is considered to be broken when ${s(1)_i}$ reaches its maximum value of $\ln 4 \approx 1.386$. 
Hence, a quantum entanglement analysis allows us to resolve bond-breaking processes of individual $\sigma$-, $\pi$-, etc., bonds in multi-bonded centers~\cite{entanglement_bonding_2013}.

\begin{figure}[t]
	\begin{center}
		\includegraphics[width=0.5\textwidth]{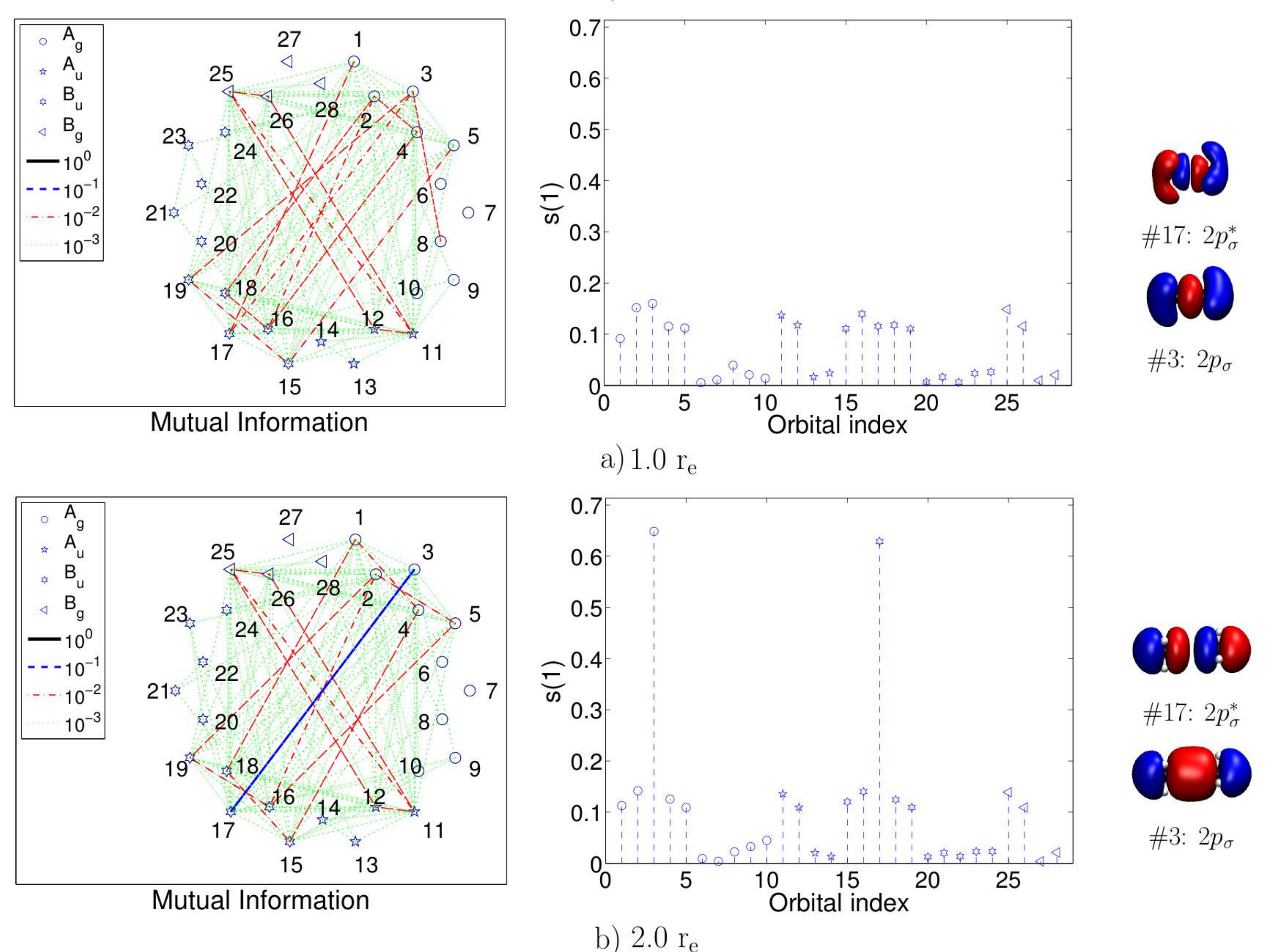}
	\end{center}
	\caption{Mutual information and single orbital entropies $s(1)_i$ for
	DMRG(14,28)$[1024,512,10^{-5}]$ calculations for the $\rm{C_2H_6}$ (${\rm r_{\rm e}}=1.53$) molecule at different inter-atomic distances. The orbitals are numbered and sorted according to their (CASSCF)
	natural occupation numbers. Strongly entangled orbitals are shown on the
	right-hand side. Each orbital index in the $s(1)_i$ diagram indicates one
	molecular orbital and corresponds to the same natural orbital as numbered in
	the mutual information plot (starting of index 1 and proceeding clockwise).}
	\label{fig:C2H6}
\end{figure}
\section{Results}
%
%%%%%%%%%%%%%%%%%%%%%%%%%%%%%%%%%%%%%%%%%%%%%%%%%%%%%%%%%%%%%%%%%%%%%%%%%%%%%%%%%%%%%%%%%%%%%%%%%%%%%%%%%%%%%%%
%%%%%%%%%%%%%%%%%%%%%%%%%%%%%%%%%%%%%%%%%%%%%%%%%%%%%%%%%%%%%%%%%%%%%%%%%%%%%%%%%%%%%%%%%%%%%%%%%%%%%%%%%%%%%%%
%%%%%%%%%%%%%%%%%%%%%%%%%%%%%%%%%%%%%%%%%%%%%%%%%%%%%%%%%%%%%%%%%%%%%%%%%%%%%%%%%%%%%%%%%%%%%%%%%%%%%%%%%%%%%%%
\subsection{Bonding between two carbon centers}
%%%%%%%%%%%%%%%%%%%%%%%%%%%%%%%%%%%%%%%%%%%%%%%%%%%%%%%%%%%%%%%%%%%%%%%%%%%%%%%%%%%%%%%%%%%%%%%%%%%%%%%%%%%%%%%
%%%%%%%%%%%%%%%%%%%%%%%%%%%%%%%%%%%%%%%%%%%%%%%%%%%%%%%%%%%%%%%%%%%%%%%%%%%%%%%%%%%%%%%%%%%%%%%%%%%%%%%%%%%%%%%
%%%%%%%%%%%%%%%%%%%%%%%%%%%%%%%%%%%%%%%%%%%%%%%%%%%%%%%%%%%%%%%%%%%%%%%%%%%%%%%%%%%%%%%%%%%%%%%%%%%%%%%%%%%%%%%
\subsubsection{Assessing the accuracy of entanglement-based bond orders for polyatomic molecules.}
%%%%%%%%%%%%%%%%%%%%%%%%%%%%%%%%%%%%%%%%%%%%%%%%%%%%%%%%%%%%%%%%%%%%%%%%%%%%%%%%%%%%%%%%%%%%%%%%%%%%%%%%%%%%%%%
%%%%%%%%%%%%%%%%%%%%%%%%%%%%%%%%%%%%%%%%%%%%%%%%%%%%%%%%%%%%%%%%%%%%%%%%%%%%%%%%%%%%%%%%%%%%%%%%%%%%%%%%%%%%%%%
\ The carbon--carbon bond is a fairly strong covalent bond, which gives rise to an enormous variety of organic molecules.
The strength of this bond largely depends on its multiplicity; the higher the multiplicity the stronger the carbon--carbon bond.
\begin{figure}[t]
	\begin{center}
		\includegraphics[width=0.5\textwidth]{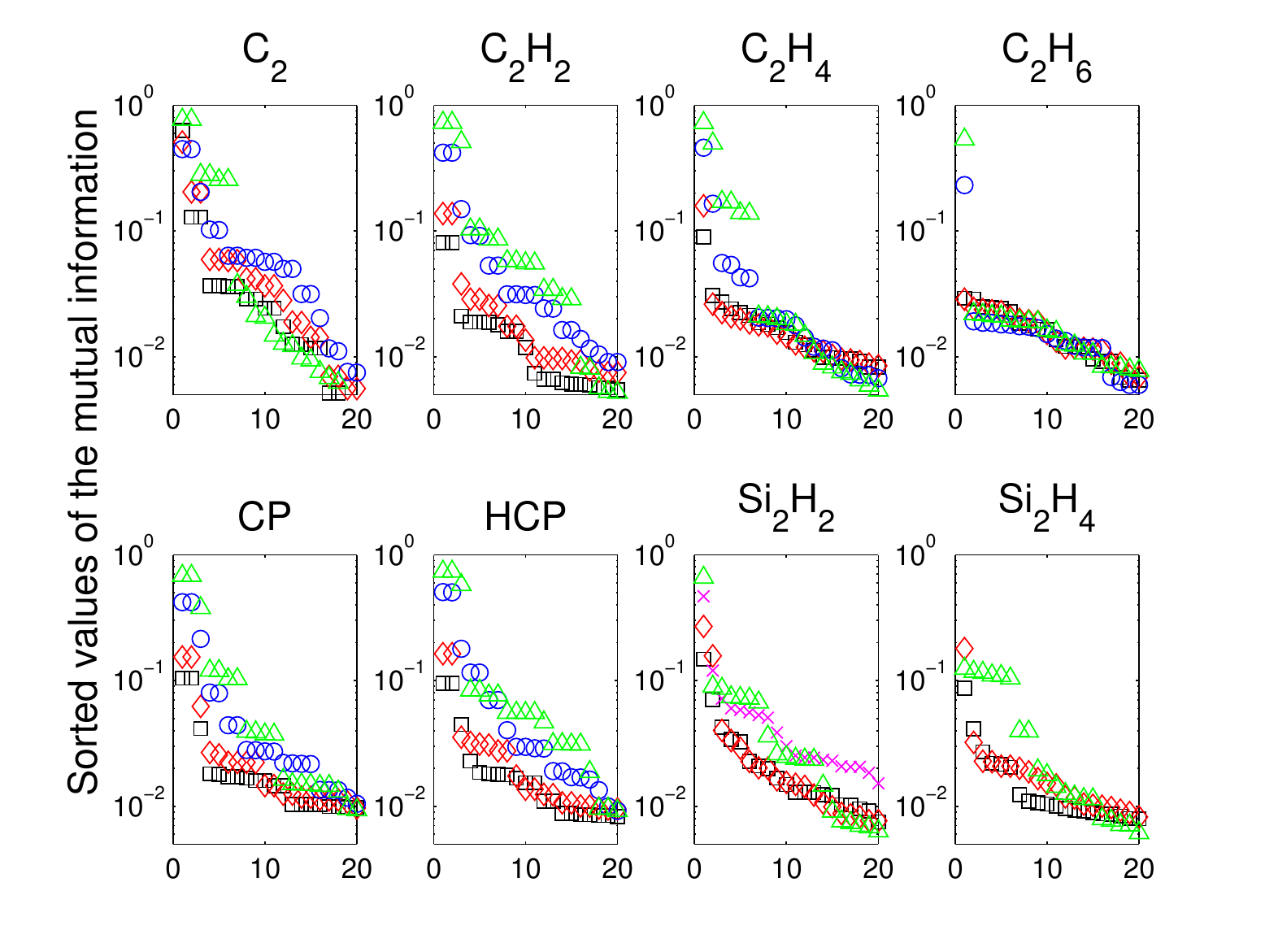}
	\end{center}
	\caption{Decay of the largest values of the mutual information ($I_{i,j}>0.005$) for all molecules and inter-atomic distances investigated in this work. 
	The symbols  {\color{black}{$\Box$}}, {\color{red}{$\Diamond$}}, {\color{blue}{$\ocircle$}}, {\color{magenta}{$\times$}} and {\color{green0}{$\vartriangle$}} stand for r=0.8, 1$\rm r_e$, 1.5$\rm r_e$, 1.65$\rm r_e$, 2$\rm r_e$, respectively.}
	\label{fig:decay_Iij}
\end{figure}
It is well-known and generally accepted that the carbon--carbon bond is weakest in alkanes, stronger in alkenes, and strongest in alkynes, where singly-, doubly- and triply- bonded centers are formed, respectively~\cite{carbon_book}.
In this study, we choose the three lightest carbon-containing representatives of the hydrocarbon series ethane (${\rm C_2H_6}$), ethene (${\rm C_2H_4}$), and acetylene (${\rm C_2H_2}$).
These molecules represent a prototypical test set to study the bond multiplicities in polyatomic molecules by means of quantum entanglement~\cite{entanglement_bonding_2013}. 

Figure~\ref{fig:C2H6} shows the mutual information and single-orbital entropies obtained from DMRG(14,28) calculations for the ${\rm C_2H_6}$ molecule at two different carbon--carbon distances: 1.0$\rm {r_e}$ and 2.0$\rm {r_e}$ (additional entanglement diagrams for 0.8$\rm {r_e}$ and 1.5$\rm {r_e}$ can be found in the Supporting Information). 
At the inter-nuclear distance of 1.0${\rm r_e}$, all orbitals are moderately and weakly entangled and it remains rather difficult to identify the bonding and antibonding combinations of molecular orbitals corresponding to the single-bond of interest. 
When the ${\rm H_3C}$ fragments are pulled apart, a gradual increase in mutual information and single-orbital entropies can be observed for the bonding and antibonding ${p_{\sigma}}$-orbitals (\#3 and \#17) in Figure~\ref{fig:C2H6} reaching the value of about 0.7 in the vicinity of dissociation (see also Figure 1 of the Supporting Information for more details). We should note that jumps in the mutual information can be indeed observed that occur around $I_{ij} \approx 10^{-1}$ (strongly entangled orbitals) and $I_{ij} \approx 10^{-2}$ (moderately entangled orbitals). Figure~\ref{fig:decay_Iij} shows the the decay of the 20 largest contributions to $I_{ij}$ for all investigated molecules and inter-atomic distances.
Our entanglement analysis correctly predicts a bond order of 1 between the two carbon centers of C$_2$H$_6$ (cf. Figure~\ref{fig:C2H6}).

\begin{figure}[t]
	\begin{center}
		\includegraphics[width=0.5\textwidth]{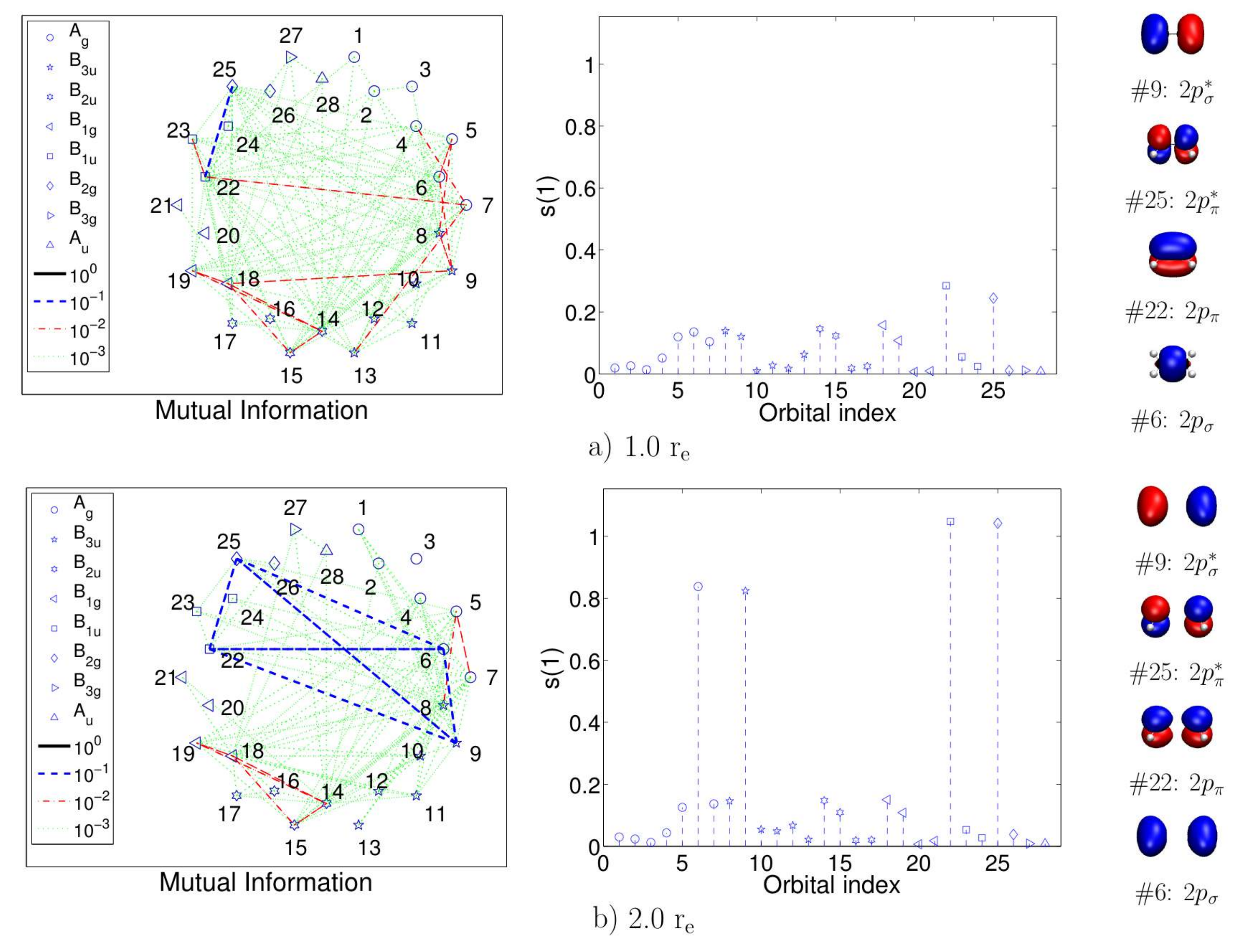}
	\end{center}
	\caption{Mutual information and single orbital entropies $s(1)_i$ for
	DMRG(12,28)$[1024,512,10^{-5}]$ calculations for the $\rm{C_2H_4}$ (${\rm r_{\rm e}}=1.34$) molecule at different inter-atomic distances. The orbitals are numbered and sorted according to their (CASSCF)
	natural occupation numbers. Strongly entangled orbitals are shown on the
	right-hand side. Each orbital index in the $s(1)_i$ diagram indicates one
	molecular orbital and corresponds to the same natural orbital as numbered in
	the mutual information plot (starting of index 1 and proceeding clockwise).}
	\label{fig:C2H4}
\end{figure}

The quantum entanglement study of the ${\rm C_2H_4}$ molecule determined from DMRG(12,28) wave functions is illustrated in Figure~\ref{fig:C2H4} (and Figure 2 of the Supporting Information). 
We obtain a similar entanglement profile as for C$_2$H$_6$, though the bonding and antibonding ${p_{\pi}}$-orbitals 
are now more strongly entangled around the equilibrium structure than the bonding and antibonding ${p_{\sigma}}$-orbitals. 
Pulling the CH$_2$ fragments apart, weakens first the $\pi$-bond followed by the (stronger) $\sigma$-bond (see Figure 2 of the Supporting Information for a complete picture). 
The entanglement-based bond order of carbon--carbon in the ${\rm C_2H_4}$ molecule is equal to 2. 

\begin{figure}[t]
	\begin{center}
       	\includegraphics[width=0.5\textwidth]{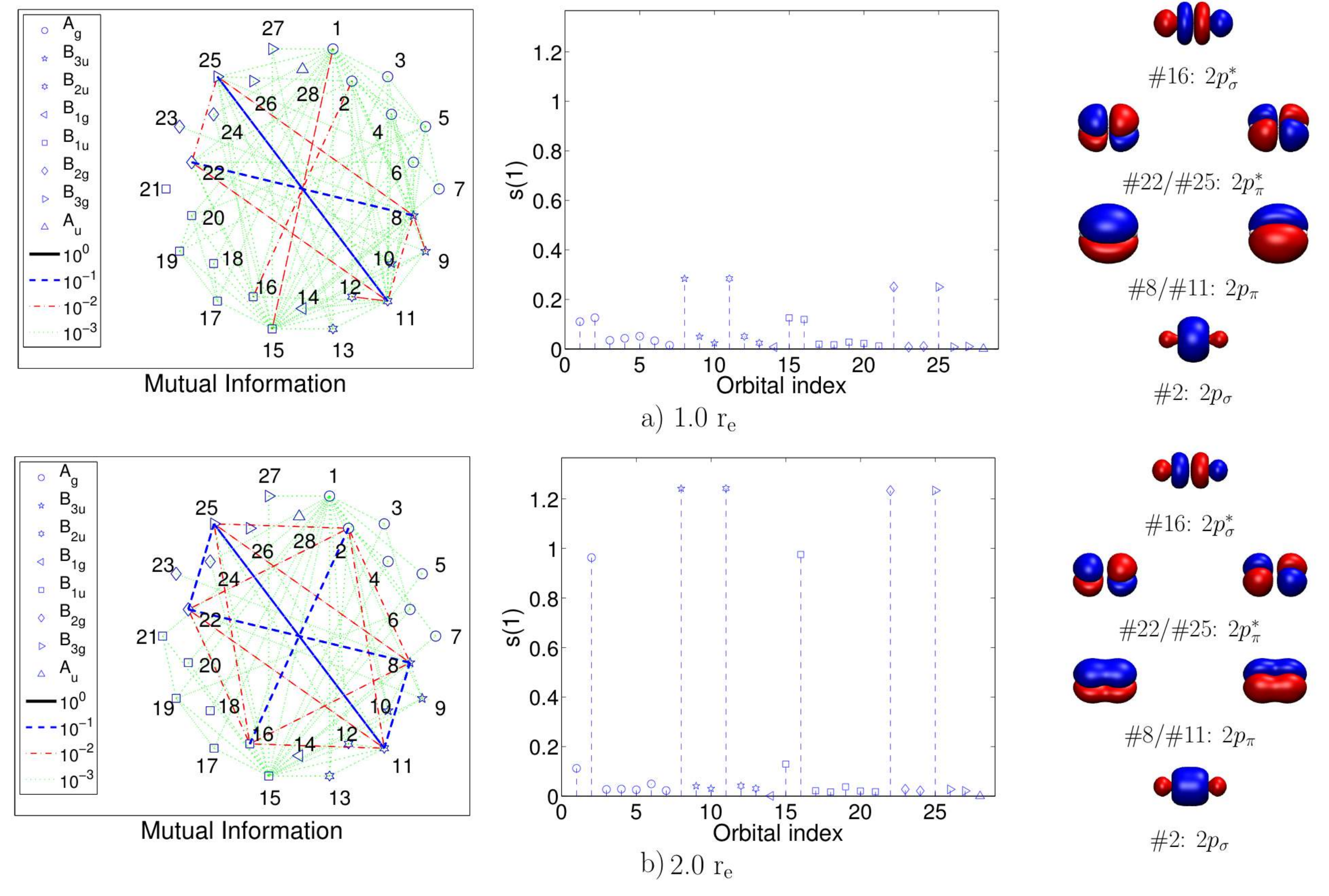}
	\end{center}
	\caption{Mutual information and single orbital entropies $s(1)_i$ for
	DMRG(10,28)$[1024,512,10^{-5}]$ calculations for the $\rm{C_2H_2}$ (${\rm r_{\rm e}}=1.21$) molecule at different inter-atomic distances. The orbitals are numbered and sorted according to their (CASSCF)
	natural occupation numbers. Strongly entangled orbitals are shown on the
	right-hand side. Each orbital index in the $s(1)_i$ diagram indicates one
	molecular orbital and corresponds to the same natural orbital as numbered in
	the mutual information plot (starting of index 1 and proceeding clockwise).}
	\label{fig:C2H2}
\end{figure}

The carbon--carbon bond in ${\rm C_2H_2}$ is dissected in Figure~\ref{fig:C2H2}. 
Similar to ${\rm C_2H_4}$, four orbitals ($\#8, \#11, \#22$ and $\#25$) have slightly higher values for the single-orbital entropy than all remaining orbitals ($s(1)_i=0.2$ vs. $s(1)_i=0.1$) around the equilibrium distance. 
These are the doubly degenerate C ${2p_{\pi}}$- and C ${2p_{\pi^*}}$-orbitals. 
Stretching the CH fragments increases, as expected, first the orbital entanglement of the C ${2p_{\pi}}$/${2p_{\pi^*}}$-orbitals, followed by the bonding and antibonding combination of the C ${2p_z}$ atomic orbitals. 
The entanglement diagram at 2.0$\rm {r_e}$ in the bottom panel of Figure~\ref{fig:C2H2} correctly predicts a triple bond between the two carbon centers in the ${\rm C_2H_2}$ molecule. 
 
In summary, our entanglement-based approach for determining bond orders, originally introduced for diatomic molecules, is transferable to polyatomic molecules and correctly predicts single, double and triple bond orders between the carbon--carbon centers in ethane, ethene and acetylene, respectively.

%%%%%%%%%%%%%%%%%%%%%%%%%%%%%%%%%%%%%%%%%%%%%%%%%%%%%%%%%%%%%%%%%%%%%%%%%%%%%%%%%%%%%%%%%%%%%%%%%%%%%%%%%%%%%%%
%%%%%%%%%%%%%%%%%%%%%%%%%%%%%%%%%%%%%%%%%%%%%%%%%%%%%%%%%%%%%%%%%%%%%%%%%%%%%%%%%%%%%%%%%%%%%%%%%%%%%%%%%%%%%%%
\subsubsection{Notorious case of bonding in C$_2$. }
%%%%%%%%%%%%%%%%%%%%%%%%%%%%%%%%%%%%%%%%%%%%%%%%%%%%%%%%%%%%%%%%%%%%%%%%%%%%%%%%%%%%%%%%%%%%%%%%%%%%%%%%%%%%%%%
%%%%%%%%%%%%%%%%%%%%%%%%%%%%%%%%%%%%%%%%%%%%%%%%%%%%%%%%%%%%%%%%%%%%%%%%%%%%%%%%%%%%%%%%%%%%%%%%%%%%%%%%%%%%%%%
The formal bond order of C$_2$ remains a challenge for present-day chemistry. 
While multireference configuration interaction calculations on the carbon dimer suggested a bond order of 2 (cf. Ref.~\citenum{Sherill_C2}), a bond order of 4 has been anticipated, as well~\cite{Shaik2012,Danovich2013,Frenking2013,Matxain2013}. 
However, the presence of a quadruple bond in C$_2$ has been discarded in the most recent Generalized Valence-Bond (GVB) study presented in Ref.~\citenum{Dunning_C2}. 
Taking into account the fact that an accurate quantum chemical description of the C$_2$ molecule requires a multireference method~\cite{Bauschlicher1987,Abrams-C2-2004,Sherill_C2,Piecuch-C2-2005,evangelista2011,Markus-chemistry,C2-Mol-Phys,Dunning_C2,CheMPS2,Gordon_2014,C2-2014}, a DMRG study combined with a quantum entanglement analysis can provide a complementary entropic perspective on the nature of bonding in C$_2$. 
\begin{figure}[t]
	\begin{center}
       	\includegraphics[width=0.5\textwidth]{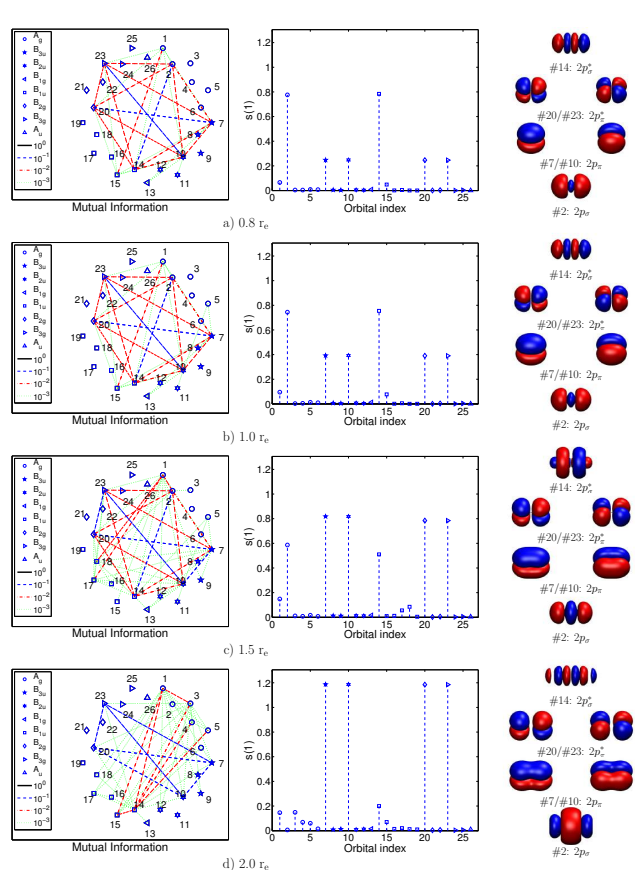}
	\end{center}
	\caption{Mutual information and single orbital entropies $s(1)_i$ for
	DMRG(8,26)$[2048,512,10^{-5}]$ calculations for the $\rm{C_2}$ (${\rm r_{\rm e}}=1.24$) molecule at different inter-atomic
	distances. The orbitals are numbered and sorted according to their (CASSCF)
	natural occupation numbers. Strongly entangled orbitals are shown on the
	right-hand side. Each orbital index in the $s(1)_i$ diagram indicates one
	molecular orbital and corresponds to the same natural orbital as numbered in
	the mutual information plot (starting of index 1 and proceeding clockwise).}
	\label{fig:C2}
\end{figure}

As summarized in Figure~\ref{fig:C2}, the C$_2$ molecule contains strongly entangled orbitals ($\#$2 and $\#$14) already around the equilibrium distance (compare Figure~\ref{fig:C2} and Figures~\ref{fig:C2H6}--\ref{fig:C2H2}).
These strongly correlated orbitals correspond to the $2p_\sigma$ and $2p_{\sigma^*}$ molecular orbitals. 
When the two carbon atoms are pulled apart, the single-orbital entropies associated with the $2p_\sigma$/$2p_{\sigma^*}$-orbital pair ($\#$2 and $\#$14) gradually decrease from 0.8 at 0.8${\rm r_e}$ to less than 0.2 at 2.0${\rm r_e}$. 
Simultaneously, the entanglement of the degenerate $2p_\pi$ and $2p_{\pi^*}$ molecular orbital pairs increases from 0.3 at 0.8${\rm r_e}$ to 1.2 at 2.0${\rm r_e}$ .   

We should note that, based on our observations, a true $\pi$-bond is associated with strongly entangled $p_{\pi}$/$p_{\pi^*}$-orbital pairs, which comprises larger $s(1)_i$-values than those obtained for $\sigma$/$\sigma^*$-orbital pairs.  
A separation of the carbon--carbon centers to 2.0${\rm r_e}$ gives rise to the traditional $\pi$-double bond (dominant contributions to $s(1)_i$) similar to that found in acetylene, while the $2p_\sigma$/$2p_{\sigma^*}$-orbital pairs do not participate in the chemical bonding (vanishing $s(1)_i$). 
Around the bond-breaking region of 1.5${\rm r_e}$, we observe a gradual transition of the singly-bonded C$_2$ molecule, over two triply-bonded centers to a purely doubly-bonded system.  

Our observations underline the complex nature of the chemical bond in the $\rm{C_2}$ molecule, which clearly deviates from the classical bonding prototype of single, double and triple bond. 
Finally, it is important to note that our entanglement analysis is in agreement with the most recent GVB study~\cite{Dunning_C2};
elucidating that the C$_2$ molecule around the equilibrium distance is well-described "as having a traditional covalent $\sigma$-bond". 
However, our entanglement analysis cannot resolve the antiferrmagnetically coupled electrons in the remaining $\pi$-orbitals~\cite{Markus-chemistry,Dunning_C2}, due to delocalized natural orbital basis used in this study, where all states equally contribute to $s(1)_i$ (see Figure 7 of the Supporting Information, where the contribution of each component of the single-orbital entropy is presented).

\begin{figure}[t]
	\begin{center}
       	\includegraphics[width=0.5\textwidth]{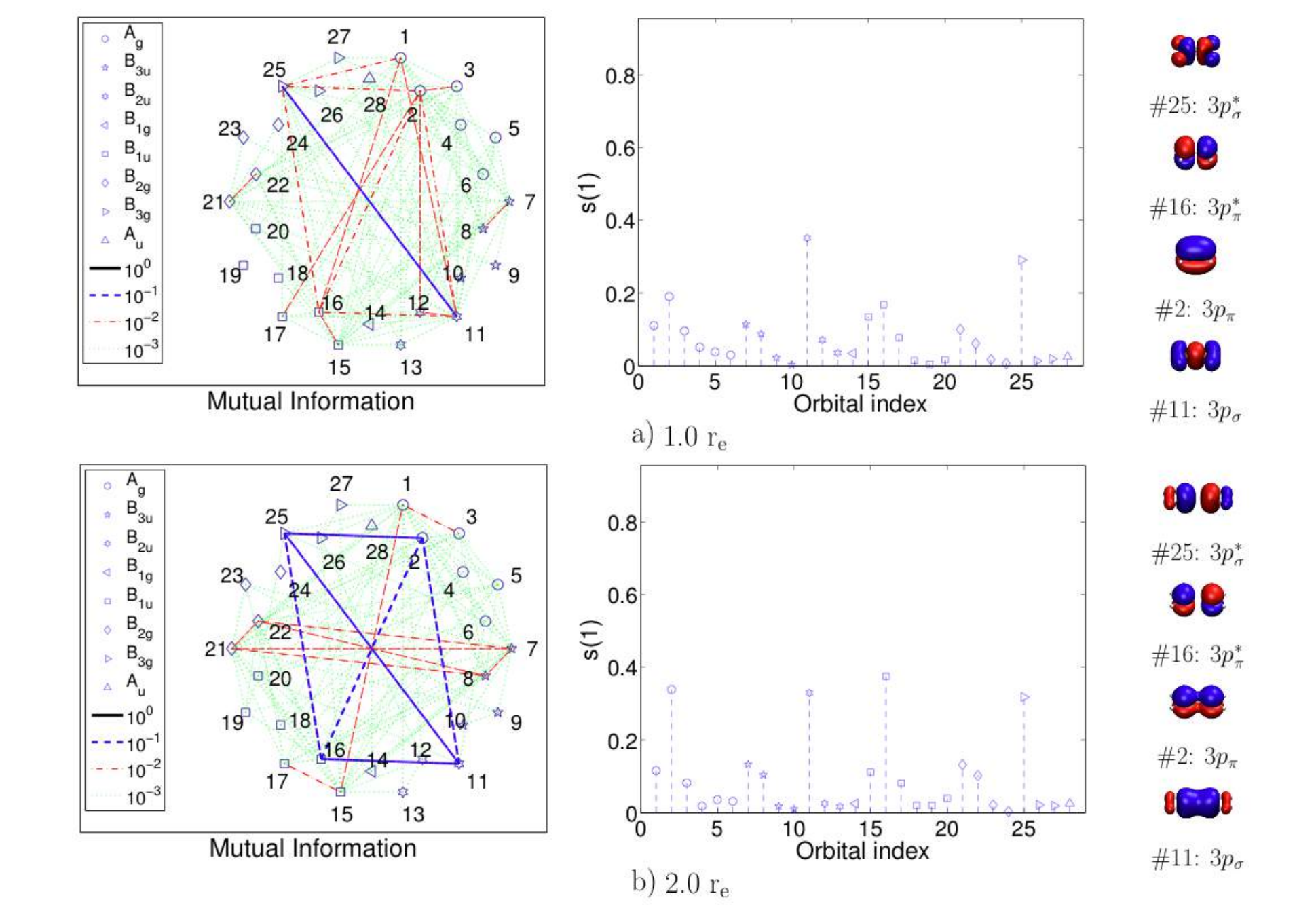}
	\end{center}
	\caption{Mutual information and single orbital entropies $s(1)_i$ for
	DMRG(12,28)$[1024,512,10^{-5}]$ calculations for the $\rm{Si_2H_4}$ (${\rm r_{\rm e}}=2.15$) molecule at different inter-atomic distances. The orbitals are numbered and sorted according to their (CASSCF)
	natural occupation numbers. Strongly entangled orbitals are shown on the
	right-hand side. Each orbital index in the $s(1)_i$ diagram indicates one
	molecular orbital and corresponds to the same natural orbital as numbered in
	the mutual information plot (starting of index 1 and proceeding clockwise).}
	\label{fig:Si2H4}
\end{figure}

\begin{figure}[h!]
	\begin{center}
       	\includegraphics[width=0.5\textwidth]{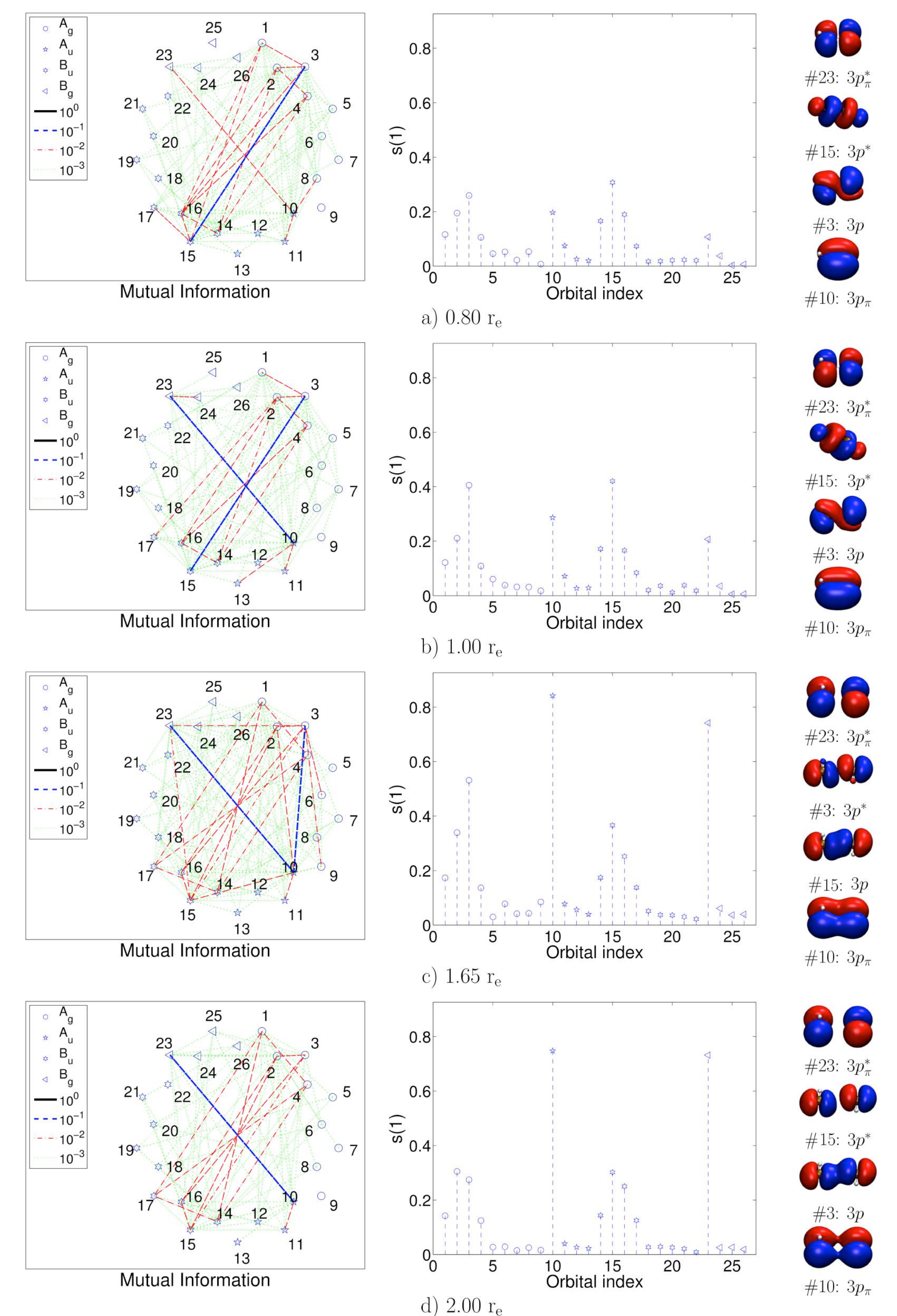}
	\end{center}
	\caption{Mutual information and single orbital entropies $s(1)_i$ for
	DMRG(10,26)$[1024,512,10^{-5}]$ calculations for the $\rm{Si_2H_2}$ (${\rm r_{\rm e}}=2.11$) molecule at different inter-atomic distances. The orbitals are numbered and sorted according to their (CASSCF)
	natural occupation numbers. Strongly entangled orbitals are shown on the
	right-hand side. Each orbital index in the $s(1)_i$ diagram indicates one
	molecular orbital and corresponds to the same natural orbital as numbered in
	the mutual information plot (starting of index 1 and proceeding clockwise).}
	\label{fig:Si2H2}
\end{figure}

\begin{figure}[b]
	\begin{center}
       	\includegraphics[width=0.5\textwidth]{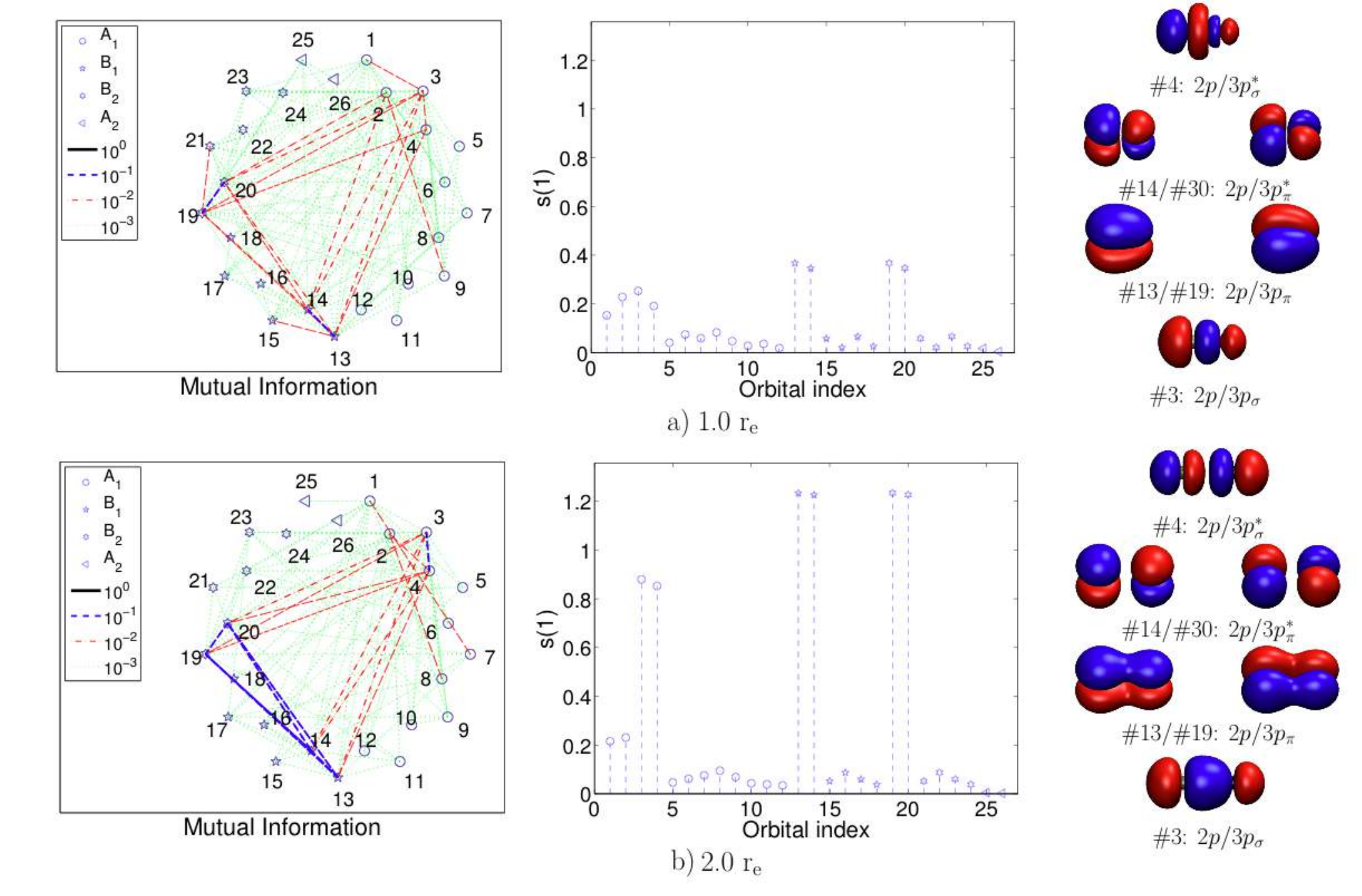}
	\end{center}
	\caption{Mutual information and single orbital entropies $s(1)_i$ for
	DMRG(10,26)$[1024,512,10^{-5}]$ calculations for the $\rm{[CP]^-}$ (${\rm r_{\rm e}}=1.62$) molecule at different inter-atomic distances. The orbitals are numbered and sorted according to their (CASSCF)
	natural occupation numbers. Strongly entangled orbitals are shown on the
	right-hand side. Each orbital index in the $s(1)_i$ diagram indicates one
	molecular orbital and corresponds to the same natural orbital as numbered in
	the mutual information plot (starting of index 1 and proceeding clockwise).}
	\label{fig:-CP}
\end{figure}
%%%%%%%%%%%%%%%%%%%%%%%%%%%%%%%%%%%%%%%%%%%%%%%%%%%%%%%%%%%%%%%%%%%%%%%%%%%%%%%%%%%%%%%%%%%%%%%%%%%%%%%%%%%%%%%
%%%%%%%%%%%%%%%%%%%%%%%%%%%%%%%%%%%%%%%%%%%%%%%%%%%%%%%%%%%%%%%%%%%%%%%%%%%%%%%%%%%%%%%%%%%%%%%%%%%%%%%%%%%%%%%
%%%%%%%%%%%%%%%%%%%%%%%%%%%%%%%%%%%%%%%%%%%%%%%%%%%%%%%%%%%%%%%%%%%%%%%%%%%%%%%%%%%%%%%%%%%%%%%%%%%%%%%%%%%%%%%
\section{Entanglement in silicon-silicon bonding}
%%%%%%%%%%%%%%%%%%%%%%%%%%%%%%%%%%%%%%%%%%%%%%%%%%%%%%%%%%%%%%%%%%%%%%%%%%%%%%%%%%%%%%%%%%%%%%%%%%%%%%%%%%%%%%%
%%%%%%%%%%%%%%%%%%%%%%%%%%%%%%%%%%%%%%%%%%%%%%%%%%%%%%%%%%%%%%%%%%%%%%%%%%%%%%%%%%%%%%%%%%%%%%%%%%%%%%%%%%%%%%%
%%%%%%%%%%%%%%%%%%%%%%%%%%%%%%%%%%%%%%%%%%%%%%%%%%%%%%%%%%%%%%%%%%%%%%%%%%%%%%%%%%%%%%%%%%%%%%%%%%%%%%%%%%%%%%%
$\rm{Si_2H_2}$ and $\rm{Si_2H_4}$ are heavier homologues of ethene and acetylene. 
They exhibit a peculiar nature of the chemical bond that is notably different from those between the carbon centers in $\rm{C_2H_2}$ and $\rm{C_2H_4}$, respectively~\cite{Lester_Si2H2}. 
The complexity of the chemical bond in these systems originates (a) from their "trans-bent" geometries~\cite{Si2H4-1,Si2H4-2,Si2H4-3} (by contrast, $\rm{C_2H_4}$ is planar and $\rm{C_2H_2}$ is linear in their equilibrium structures), (b) from the energetical preference for low-spin states located at the Si centers upon Si--Si bond dissociation~\cite{Si2H2_PRL,Lein2005,Si2H2-good,Adamczyk2011}, and (c) differences in the spatial distribution of $s$- and $p$-orbitals~\cite{kutzelnigg_bonding1}.

For simplicity, we imposed a planar geometry on the Si$_2$H$_4$ molecule.  
Figure~\ref{fig:Si2H4} illustrates the performance of the entanglement-based bond analysis for $\rm{Si_2H_4}$ obtained from DMRG(12,28) calculations. 
The entanglement diagrams are similar to those of C$_2$H$_4$ indicating doubly-bonded Si--Si centers. 
The double bond in the Si$_2$H$_4$ compound in its planar structure was also approved in recent theoretical studies~\cite{Si2H4p1,Si2H4p2,Si2H4p3,Si2H4p4,Si2H4p5}.  

Figure~\ref{fig:Si2H2} depicts the entanglement profiles for breaking the Si--Si centers in the Si$_2$H$_2$ molecule, obtained form DMRG(10,28) calculations, where we imposed the "trans-bent" geometry close to the equilibrium.  
At 0.8${\rm r_e}$ and 1.0${\rm r_e}$, four orbitals are moderately entangled ($\#$3, $\#$10, $\#$15, $\#$23) that correspond to the 3${p}_{n}$- and 3${p}_{n^*}$-orbitals (\#3 and \#10, lone-pair type bonds~\cite{Lein2005}) and to the 3${p}_{\pi}$- and 3${p}_{\pi^*}$-orbitals (\#16 and \#23), respectively.
Stretching the Si--Si centers leads to structural changes from "trans-bent" to "trans-perpendicular" geometries (see Table VI of the Supporting Information for more details).   
Upon dissociation, the single-orbital entropy of orbitals involved in $\pi$-type bonding increases fastest and reaches a maximum value of 0.85 at 1.5${\rm r_e}$. 
At larger inter-atomic Si--Si distances, $s(1)_i$ slightly decreases to 0.8, while the single-orbital entropies of the lone-pair orbital pairs decline to about 0.3 around 2.0${\rm r_e}$. 
This peculiar behavior is caused by structural changes and the reorientation of the Si--H fragments with respect to each other upon dissociation. 
The quantum-entanglement diagrams extracted at 2.0${\rm r_e}$ point to a single $\pi$-type bonding in the $\rm{Si_2H_2}$ molecule, supported by one weakly entangled bond attributed to the interaction between both Si lone-pairs. 
We should note, however, that this lone-pair 'bond'~\cite{Lein2005} comprises the most strongly entangled orbitals around the equilibrium structure. 
 
The entanglement-based bond analysis agrees well with other theoretical findings that the Si$_2$H$_2$ molecule has a double bond in its "trans-bent" geometry and a single bond in its "trans-perpendicular" structure~\cite{Lein2005,ABC_bonding,Adamczyk2011}.
Our study also supports that the Si--Si bond in Si$_2$H$_2$ is a combination of one $\pi$- and one "donor--acceptor" lone-pair-bond~\cite{Lein2005,Adamczyk2011}.

\begin{figure}[b]
	\begin{center}
       	\includegraphics[width=0.5\textwidth]{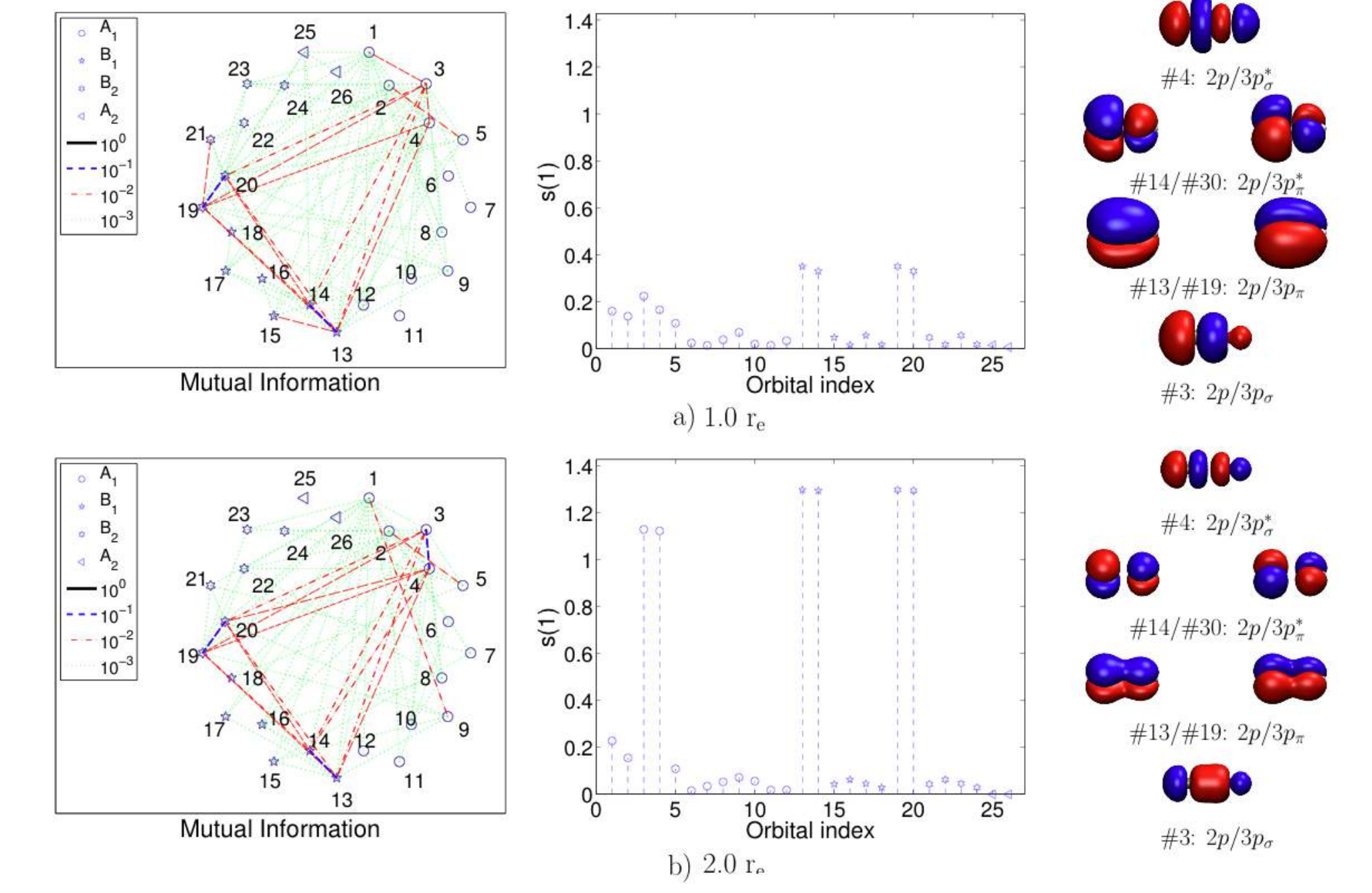}
	\end{center}
	\caption{Mutual information and single orbital entropies $s(1)_i$ for
	DMRG(14,28)$[1024,512,10^{-5}]$ calculations for the $\rm{HCP}$ (${\rm r_{\rm e}}=1.55$) molecule at different inter-atomic distances. The orbitals are numbered and sorted according to their (CASSCF)
	natural occupation numbers. Strongly entangled orbitals are shown on the
	right-hand side. Each orbital index in the $s(1)_i$ diagram indicates one
	molecular orbital and corresponds to the same natural orbital as numbered in
	the mutual information plot (starting of index 1 and proceeding clockwise).}
	\label{fig:HCP}
\end{figure}
%
%%%%%%%%%%%%%%%%%%%%%%%%%%%%%%%%%%%%%%%%%%%%%%%%%%%%%%%%%%%%%%%%%%%%%%%%%%%%%%%%%%%%%%%%%%%%%%%%%%%%%%%%%%%%%%%
%%%%%%%%%%%%%%%%%%%%%%%%%%%%%%%%%%%%%%%%%%%%%%%%%%%%%%%%%%%%%%%%%%%%%%%%%%%%%%%%%%%%%%%%%%%%%%%%%%%%%%%%%%%%%%%
%%%%%%%%%%%%%%%%%%%%%%%%%%%%%%%%%%%%%%%%%%%%%%%%%%%%%%%%%%%%%%%%%%%%%%%%%%%%%%%%%%%%%%%%%%%%%%%%%%%%%%%%%%%%%%%
\section{Bond multiplicity in carbon--phosphorus centers}
%%%%%%%%%%%%%%%%%%%%%%%%%%%%%%%%%%%%%%%%%%%%%%%%%%%%%%%%%%%%%%%%%%%%%%%%%%%%%%%%%%%%%%%%%%%%%%%%%%%%%%%%%%%%%%%
%%%%%%%%%%%%%%%%%%%%%%%%%%%%%%%%%%%%%%%%%%%%%%%%%%%%%%%%%%%%%%%%%%%%%%%%%%%%%%%%%%%%%%%%%%%%%%%%%%%%%%%%%%%%%%%
%%%%%%%%%%%%%%%%%%%%%%%%%%%%%%%%%%%%%%%%%%%%%%%%%%%%%%%%%%%%%%%%%%%%%%%%%%%%%%%%%%%%%%%%%%%%%%%%%%%%%%%%%%%%%%%

The discovery of phosphoalkynes by Gier~\cite{Gier_1961} in 1961 paved the way the way for a new type of chemistry. 
The more diffuse nature of the phosphorus valence orbitals gives rise to a variety of interesting bonding properties with lighter first-row elements.
Notably, the ground state of the phosphorus atom is a quadruplet, allowing for a formal triple bond with the carbon atom. 

This hypothesis can be elucidated by studying the quantum entanglement of carbon--phosphorus bonding in the CP and [HCP$^-$] prototypical molecules. 
An entanglement-bond analysis based on DMRG(10,26) calculations is presented in Figures~\ref{fig:-CP} and~\ref{fig:HCP} for 
the CP and [HCP$^-$] moieties, respectively.
At a distance of 1.0$\rm{r_e}$ in both molecules, the $\pi$/$\pi^*$-orbitals (\#14, \#13, \#19, and \#30) are moderately entangled while all other orbitals are only weakly entangled.
Upon dissociation the single-orbital entropies associated with the degenerate $\pi$/$\pi^*$-orbital pairs, corresponding to the bonding and antibonding combinations of C 2${p}_{\pi}$- and P 3${p}_{\pi}$-orbitals, increase faster than those of the $\sigma$/$\sigma^*$-orbital pairs, ($\#$3 and $\#$4) which correspond to the bonding and antibonding combinations of C 2${p}_{\sigma}$- and P 3${p}_{\sigma}$-orbitals. 
In both molecules, the maximum of $s(1)_i$ for $\pi$-orbitals is reached at 2.0$\rm r_e$ and equals approximately 1.3. 
Additional presence of highly entangled $\sigma$-orbitals at this distance clearly indicates a triple bond between carbon and phosphorus in the CP and [HCP$^-$] molecules. 
The entanglement nature of this bond is quite similar to what we observed for the triply bonded carbon--carbon centers in the ${\rm C_2H_2}$ molecule (cf. Figure~\ref{fig:C2H2}) as well as to the triply bonded dinitrogen molecule discussed in Ref.~\citenum{entanglement_bonding_2013}.  
We should emphasize that all triply bonded centers studied possess similar decays of the mutual information as illustrated in Figure~\ref{fig:decay_Iij}. 
Finally, it is worth to mention that our findings are in line with previously reported density functional theory studies of carbon--phosphorus bonding in the $[$CP$]^-$ and HCP molecules~\cite{Lucas2008}. 

%%%%%%%%%%%%%%%%%%%%%%%%%%%%%%%%%%%%%%%%%%%%%%%%%%%%%%%%%%%%%%%%%%%%%%%%%%%%%%%%%%%%%%%%%%%%%%%%%%%%%%%%%%%%%%%
%%%%%%%%%%%%%%%%%%%%%%%%%%%%%%%%%%%%%%%%%%%%%%%%%%%%%%%%%%%%%%%%%%%%%%%%%%%%%%%%%%%%%%%%%%%%%%%%%%%%%%%%%%%%%%%
%%%%%%%%%%%%%%%%%%%%%%%%%%%%%%%%%%%%%%%%%%%%%%%%%%%%%%%%%%%%%%%%%%%%%%%%%%%%%%%%%%%%%%%%%%%%%%%%%%%%%%%%%%%%%%%
\section{Conclusions}
%%%%%%%%%%%%%%%%%%%%%%%%%%%%%%%%%%%%%%%%%%%%%%%%%%%%%%%%%%%%%%%%%%%%%%%%%%%%%%%%%%%%%%%%%%%%%%%%%%%%%%%%%%%%%%%
%%%%%%%%%%%%%%%%%%%%%%%%%%%%%%%%%%%%%%%%%%%%%%%%%%%%%%%%%%%%%%%%%%%%%%%%%%%%%%%%%%%%%%%%%%%%%%%%%%%%%%%%%%%%%%%
%%%%%%%%%%%%%%%%%%%%%%%%%%%%%%%%%%%%%%%%%%%%%%%%%%%%%%%%%%%%%%%%%%%%%%%%%%%%%%%%%%%%%%%%%%%%%%%%%%%%%%%%%%%%%%%

We presented an entanglement-based bond order analysis of carbon--carbon, silicon--silicon and carbon--phosphorus centers. 
Our approach embraces the mutual information and single-orbital entropy and correctly predicts bond multiplicities in simple polyatomic molecules like ethane, ethene and acetylene, and confirms the triple bonding between the carbon--phosphorus centers in the [CP]$^{-}$ and HCP molecules. 
The behavior of phosphorus in these prototypical phosphoalkynes closely resembles the bonding situation in the $\rm{N_2}$ and C$_2$H$_2$ molecules. 

Furthermore, our analysis confirms that the nature of the chemical bond in the C$_2$ molecule is far more complicated than for their higher substituted analogs like C$_2$H$_6$, C$_2$H$_4$, and C$_2$H$_2$. 
Stretching the C--C bond in C$_2$ does not yield a gradually increasing entanglement pattern, which is an indication of the complexity of chemical bonding in the carbon dimer. 
We also observed such an exceptional behavior of chemical bonding between the Si--Si centers in the Si$_2$H$_2$ molecule which can be explained by structural rearrangements of the SiH fragments upon dissociation.

%%%%%%%%%%%%%%%%%%%%%%%%%%%%%%%%%%%%%%%%%%%%%%%%%%%%%%%%%%%%%%%%%%%%%%%%%%%%%%%%%%%%%%%%%%%%%%%%%%%%%%%%%%%%%%%
%%%%%%%%%%%%%%%%%%%%%%%%%%%%%%%%%%%%%%%%%%%%%%%%%%%%%%%%%%%%%%%%%%%%%%%%%%%%%%%%%%%%%%%%%%%%%%%%%%%%%%%%%%%%%%%
%%%%%%%%%%%%%%%%%%%%%%%%%%%%%%%%%%%%%%%%%%%%%%%%%%%%%%%%%%%%%%%%%%%%%%%%%%%%%%%%%%%%%%%%%%%%%%%%%%%%%%%%%%%%%%%
\section{Acknowledgments}
%%%%%%%%%%%%%%%%%%%%%%%%%%%%%%%%%%%%%%%%%%%%%%%%%%%%%%%%%%%%%%%%%%%%%%%%%%%%%%%%%%%%%%%%%%%%%%%%%%%%%%%%%%%%%%%
%%%%%%%%%%%%%%%%%%%%%%%%%%%%%%%%%%%%%%%%%%%%%%%%%%%%%%%%%%%%%%%%%%%%%%%%%%%%%%%%%%%%%%%%%%%%%%%%%%%%%%%%%%%%%%%
%%%%%%%%%%%%%%%%%%%%%%%%%%%%%%%%%%%%%%%%%%%%%%%%%%%%%%%%%%%%%%%%%%%%%%%%%%%%%%%%%%%%%%%%%%%%%%%%%%%%%%%%%%%%%%%

The authors gratefully acknowledge financial support by the Swiss national science foundation SNF (project 200020-144458/1),
and from the Hungarian Research Fund (OTKA) under Grant NOs. ~K100908 and NN110360.
K.B. acknowledges the financial support from the Swiss National Science Foundation (P2EZP2 148650).
{\"O}.L.\ acknowledges support from the Alexander von Humboldt foundation and from ETH Z\"urich
during his time as a visiting professor.

%%%%%%%%%%%%%%%%%%%%%%%%%%%%%%%%%%%%%%%%%%%%%%%%%%%%%%%%%%%%%%%%%%%%%%%%%%%%%%%%%
% BIBLIOGRAPHY
%\bibliography{rsc}   % Produces the bibliography via BibTeX.

\begin{mcitethebibliography}{97}
\providecommand*{\natexlab}[1]{#1}
\providecommand*{\mciteSetBstSublistMode}[1]{}
\providecommand*{\mciteSetBstMaxWidthForm}[2]{}
\providecommand*{\mciteBstWouldAddEndPuncttrue}
  {\def\EndOfBibitem{\unskip.}}
\providecommand*{\mciteBstWouldAddEndPunctfalse}
  {\let\EndOfBibitem\relax}
\providecommand*{\mciteSetBstMidEndSepPunct}[3]{}
\providecommand*{\mciteSetBstSublistLabelBeginEnd}[3]{}
\providecommand*{\EndOfBibitem}{}
\mciteSetBstSublistMode{f}
\mciteSetBstMaxWidthForm{subitem}
{(\emph{\alph{mcitesubitemcount}})}
\mciteSetBstSublistLabelBeginEnd{\mcitemaxwidthsubitemform\space}
{\relax}{\relax}

\bibitem[Bartlett(1984)]{bartlett_book_1984}
R.~Bartlett, \emph{{Comparison of Ab Initio Quantum Chemistry with Experiment
  for Small Molecules}}, Springer, 1984\relax
\mciteBstWouldAddEndPuncttrue
\mciteSetBstMidEndSepPunct{\mcitedefaultmidpunct}
{\mcitedefaultendpunct}{\mcitedefaultseppunct}\relax
\EndOfBibitem
\bibitem[Helgaker \emph{et~al.}(2000)Helgaker, J{\o}rgensen, and
  Olsen]{Helgaker_book}
T.~Helgaker, P.~J{\o}rgensen and J.~Olsen, \emph{Molecular Electronic-Structure
  Theory}, Wiley, 2000\relax
\mciteBstWouldAddEndPuncttrue
\mciteSetBstMidEndSepPunct{\mcitedefaultmidpunct}
{\mcitedefaultendpunct}{\mcitedefaultseppunct}\relax
\EndOfBibitem
\bibitem[Bartlett and Musia{\l}(2007)]{bartlett_2007}
R.~J. Bartlett and M.~Musia{\l}, \emph{Rev.~Mod.~Phys.}, 2007, \textbf{79},
  291--350\relax
\mciteBstWouldAddEndPuncttrue
\mciteSetBstMidEndSepPunct{\mcitedefaultmidpunct}
{\mcitedefaultendpunct}{\mcitedefaultseppunct}\relax
\EndOfBibitem
\bibitem[Cohen \emph{et~al.}(2012)Cohen, Mori-S\'anchez, and
  Yang]{dft_rev_2012}
A.~J. Cohen, P.~Mori-S\'anchez and W.~Yang, \emph{Chem.~Rev.}, 2012,
  \textbf{112}, 289\relax
\mciteBstWouldAddEndPuncttrue
\mciteSetBstMidEndSepPunct{\mcitedefaultmidpunct}
{\mcitedefaultendpunct}{\mcitedefaultseppunct}\relax
\EndOfBibitem
\bibitem[Kutzelnigg(1984)]{kutzelnigg_bonding1}
W.~Kutzelnigg, \emph{Angew. Chem. Int. Ed. Engl.}, 1984, \textbf{23},
  272--295\relax
\mciteBstWouldAddEndPuncttrue
\mciteSetBstMidEndSepPunct{\mcitedefaultmidpunct}
{\mcitedefaultendpunct}{\mcitedefaultseppunct}\relax
\EndOfBibitem
\bibitem[Clark and Davidson(2001)]{Clark2001}
A.~E. Clark and E.~R. Davidson, \emph{J. Chem. Phys.}, 2001, \textbf{115},
  7382--7392\relax
\mciteBstWouldAddEndPuncttrue
\mciteSetBstMidEndSepPunct{\mcitedefaultmidpunct}
{\mcitedefaultendpunct}{\mcitedefaultseppunct}\relax
\EndOfBibitem
\bibitem[Davidson and Clark(2001)]{Davidson2001}
E.~R. Davidson and A.~E. Clark, \emph{Mol. Phys.}, 2001, \textbf{100},
  373--383\relax
\mciteBstWouldAddEndPuncttrue
\mciteSetBstMidEndSepPunct{\mcitedefaultmidpunct}
{\mcitedefaultendpunct}{\mcitedefaultseppunct}\relax
\EndOfBibitem
\bibitem[Clark and Davidson(2002)]{Clark2002}
A.~E. Clark and E.~R. Davidson, \emph{J. Phys. Chem. A}, 2002, \textbf{106},
  6890--6896\relax
\mciteBstWouldAddEndPuncttrue
\mciteSetBstMidEndSepPunct{\mcitedefaultmidpunct}
{\mcitedefaultendpunct}{\mcitedefaultseppunct}\relax
\EndOfBibitem
\bibitem[Herrmann \emph{et~al.}(2005)Herrmann, Reiher, and Hess]{Herrmann2005}
C.~Herrmann, M.~Reiher and B.~A. Hess, \emph{J. Chem. Phys.}, 2005,
  \textbf{122}, 34102\relax
\mciteBstWouldAddEndPuncttrue
\mciteSetBstMidEndSepPunct{\mcitedefaultmidpunct}
{\mcitedefaultendpunct}{\mcitedefaultseppunct}\relax
\EndOfBibitem
\bibitem[Herrmann \emph{et~al.}(2006)Herrmann, Yu, and Reiher]{markusjcc2006}
C.~Herrmann, L.~Yu and M.~Reiher, \emph{J. Comput. Chem.}, 2006, \textbf{27},
  1223--1239\relax
\mciteBstWouldAddEndPuncttrue
\mciteSetBstMidEndSepPunct{\mcitedefaultmidpunct}
{\mcitedefaultendpunct}{\mcitedefaultseppunct}\relax
\EndOfBibitem
\bibitem[Mayer(2007)]{Mayer2007}
I.~Mayer, \emph{Chem. Phys. Lett.}, 2007, \textbf{440}, 357--359\relax
\mciteBstWouldAddEndPuncttrue
\mciteSetBstMidEndSepPunct{\mcitedefaultmidpunct}
{\mcitedefaultendpunct}{\mcitedefaultseppunct}\relax
\EndOfBibitem
\bibitem[Davidson and Clark(2007)]{Davidson2007a}
E.~R. Davidson and A.~E. Clark, \emph{Phys. Chem. Chem. Phys.}, 2007,
  \textbf{9}, 1881--1894\relax
\mciteBstWouldAddEndPuncttrue
\mciteSetBstMidEndSepPunct{\mcitedefaultmidpunct}
{\mcitedefaultendpunct}{\mcitedefaultseppunct}\relax
\EndOfBibitem
\bibitem[Reiher(2007)]{markus_fd}
M.~Reiher, \emph{Faraday Discuss.}, 2007, \textbf{135}, 97--124\relax
\mciteBstWouldAddEndPuncttrue
\mciteSetBstMidEndSepPunct{\mcitedefaultmidpunct}
{\mcitedefaultendpunct}{\mcitedefaultseppunct}\relax
\EndOfBibitem
\bibitem[Podewitz \emph{et~al.}(2008)Podewitz, Herrmann, Malassa, Westerhausen,
  and Reiher]{Podewitz2008}
M.~Podewitz, C.~Herrmann, A.~Malassa, M.~Westerhausen and M.~Reiher,
  \emph{Chem. Phys. Lett.}, 2008, \textbf{451}, 301--308\relax
\mciteBstWouldAddEndPuncttrue
\mciteSetBstMidEndSepPunct{\mcitedefaultmidpunct}
{\mcitedefaultendpunct}{\mcitedefaultseppunct}\relax
\EndOfBibitem
\bibitem[Alcoba \emph{et~al.}(2009)Alcoba, Lain, Torre, and
  Bochicchio]{Alcoba2009}
D.~R. Alcoba, L.~Lain, A.~Torre and R.~C. Bochicchio, \emph{Chem. Phys. Lett.},
  2009, \textbf{470}, 136--139\relax
\mciteBstWouldAddEndPuncttrue
\mciteSetBstMidEndSepPunct{\mcitedefaultmidpunct}
{\mcitedefaultendpunct}{\mcitedefaultseppunct}\relax
\EndOfBibitem
\bibitem[Mayer(2009)]{Mayer2009}
I.~Mayer, \emph{Chem. Phys. Lett.}, 2009, \textbf{478}, 323--326\relax
\mciteBstWouldAddEndPuncttrue
\mciteSetBstMidEndSepPunct{\mcitedefaultmidpunct}
{\mcitedefaultendpunct}{\mcitedefaultseppunct}\relax
\EndOfBibitem
\bibitem[Podewitz and Reiher(2010)]{Podewitz2010}
M.~Podewitz and M.~Reiher, \emph{Adv. Inorg. Chem.}, 2010, \textbf{62},
  177--230\relax
\mciteBstWouldAddEndPuncttrue
\mciteSetBstMidEndSepPunct{\mcitedefaultmidpunct}
{\mcitedefaultendpunct}{\mcitedefaultseppunct}\relax
\EndOfBibitem
\bibitem[Torre \emph{et~al.}(2010)Torre, Alcoba, Lain, and
  Bochicchio]{Torre2010}
A.~Torre, D.~R. Alcoba, L.~Lain and R.~C. Bochicchio, \emph{J. Phys. Chem. A},
  2010, \textbf{114}, 2344--2349\relax
\mciteBstWouldAddEndPuncttrue
\mciteSetBstMidEndSepPunct{\mcitedefaultmidpunct}
{\mcitedefaultendpunct}{\mcitedefaultseppunct}\relax
\EndOfBibitem
\bibitem[Mayer and Matito(2010)]{Mayer2010}
I.~Mayer and E.~Matito, \emph{Phys. Chem. Chem. Phys.}, 2010, \textbf{12},
  11308--11314\relax
\mciteBstWouldAddEndPuncttrue
\mciteSetBstMidEndSepPunct{\mcitedefaultmidpunct}
{\mcitedefaultendpunct}{\mcitedefaultseppunct}\relax
\EndOfBibitem
\bibitem[Alcoba \emph{et~al.}(2011)Alcoba, Torre, Lain, and
  Bochicchio]{Alcoba2011}
D.~R. Alcoba, A.~Torre, L.~Lain and R.~C. Bochicchio, \emph{J. Chem. Theory
  Comput.}, 2011, \textbf{7}, 3560--3566\relax
\mciteBstWouldAddEndPuncttrue
\mciteSetBstMidEndSepPunct{\mcitedefaultmidpunct}
{\mcitedefaultendpunct}{\mcitedefaultseppunct}\relax
\EndOfBibitem
\bibitem[Alcoba \emph{et~al.}(2011)Alcoba, Torre, Lain, and
  Bochicchio]{Alcoba2011a}
D.~R. Alcoba, A.~Torre, L.~Lain and R.~C. Bochicchio, \emph{Chem. Phys. Lett.},
  2011, \textbf{504}, 236--240\relax
\mciteBstWouldAddEndPuncttrue
\mciteSetBstMidEndSepPunct{\mcitedefaultmidpunct}
{\mcitedefaultendpunct}{\mcitedefaultseppunct}\relax
\EndOfBibitem
\bibitem[Ramos-Cordoba \emph{et~al.}(2012)Ramos-Cordoba, Matito, Mayer, and
  Salvador]{Ramos-Cordoba2012}
E.~Ramos-Cordoba, E.~Matito, I.~Mayer and P.~Salvador, \emph{J. Chem. Theory
  Comput.}, 2012, \textbf{8}, 1270--1279\relax
\mciteBstWouldAddEndPuncttrue
\mciteSetBstMidEndSepPunct{\mcitedefaultmidpunct}
{\mcitedefaultendpunct}{\mcitedefaultseppunct}\relax
\EndOfBibitem
\bibitem[Geerlings \emph{et~al.}(2003)Geerlings, {De Proft}, and
  Langenaeker]{Geerlings_rev_2003}
P.~Geerlings, F.~{De Proft} and W.~Langenaeker, \emph{Chem. Rev.}, 2003,
  \textbf{103}, 1793--1873\relax
\mciteBstWouldAddEndPuncttrue
\mciteSetBstMidEndSepPunct{\mcitedefaultmidpunct}
{\mcitedefaultendpunct}{\mcitedefaultseppunct}\relax
\EndOfBibitem
\bibitem[Wiberg(1968)]{Wiberg1968}
K.~B. Wiberg, \emph{Tetrahedron}, 1968, \textbf{24}, 1083--1096\relax
\mciteBstWouldAddEndPuncttrue
\mciteSetBstMidEndSepPunct{\mcitedefaultmidpunct}
{\mcitedefaultendpunct}{\mcitedefaultseppunct}\relax
\EndOfBibitem
\bibitem[Armstrong \emph{et~al.}(1973)Armstrong, Perkins, and
  Stewart]{Armstrong1973}
D.~R. Armstrong, P.~G. Perkins and J.~J.~P. Stewart, \emph{J. Chem. Soc.{,}
  Dalton Trans.}, 1973,  838--840\relax
\mciteBstWouldAddEndPuncttrue
\mciteSetBstMidEndSepPunct{\mcitedefaultmidpunct}
{\mcitedefaultendpunct}{\mcitedefaultseppunct}\relax
\EndOfBibitem
\bibitem[Natiello and Medrano(1984)]{Natiello1984}
M.~A. Natiello and J.~A. Medrano, \emph{Chem. Phys. Lett.}, 1984, \textbf{105},
  180--182\relax
\mciteBstWouldAddEndPuncttrue
\mciteSetBstMidEndSepPunct{\mcitedefaultmidpunct}
{\mcitedefaultendpunct}{\mcitedefaultseppunct}\relax
\EndOfBibitem
\bibitem[Mayer(1985)]{Mayer1985}
I.~Mayer, \emph{Theor. Chim. Acta}, 1985, \textbf{67}, 315--322\relax
\mciteBstWouldAddEndPuncttrue
\mciteSetBstMidEndSepPunct{\mcitedefaultmidpunct}
{\mcitedefaultendpunct}{\mcitedefaultseppunct}\relax
\EndOfBibitem
\bibitem[Mayer(1986)]{Mayer1986}
I.~Mayer, \emph{Int. J. Quantum Chem.}, 1986, \textbf{29}, 73--84\relax
\mciteBstWouldAddEndPuncttrue
\mciteSetBstMidEndSepPunct{\mcitedefaultmidpunct}
{\mcitedefaultendpunct}{\mcitedefaultseppunct}\relax
\EndOfBibitem
\bibitem[Bochicchio(1991)]{Bochicchio1991a}
R.~C. Bochicchio, \emph{J. Mol. Struct.}, 1991, \textbf{228}, 209--225\relax
\mciteBstWouldAddEndPuncttrue
\mciteSetBstMidEndSepPunct{\mcitedefaultmidpunct}
{\mcitedefaultendpunct}{\mcitedefaultseppunct}\relax
\EndOfBibitem
\bibitem[Alcoba \emph{et~al.}(2007)Alcoba, Bochicchio, Lain, and
  Torre]{Alcoba2007}
D.~R. Alcoba, R.~C. Bochicchio, L.~Lain and A.~Torre, \emph{Chem. Phys. Lett.},
  2007, \textbf{442}, 157--163\relax
\mciteBstWouldAddEndPuncttrue
\mciteSetBstMidEndSepPunct{\mcitedefaultmidpunct}
{\mcitedefaultendpunct}{\mcitedefaultseppunct}\relax
\EndOfBibitem
\bibitem[Mayer(2007)]{Mayer2007a}
I.~Mayer, \emph{J. Comput. Chem.}, 2007, \textbf{28}, 204--221\relax
\mciteBstWouldAddEndPuncttrue
\mciteSetBstMidEndSepPunct{\mcitedefaultmidpunct}
{\mcitedefaultendpunct}{\mcitedefaultseppunct}\relax
\EndOfBibitem
\bibitem[Ramos-Cordoba \emph{et~al.}(2013)Ramos-Cordoba, Salvador, and
  Reiher]{Markus-chemistry}
E.~Ramos-Cordoba, P.~Salvador and M.~Reiher, \emph{Chem. Eur. J.}, 2013,
  \textbf{19}, 15267--15275\relax
\mciteBstWouldAddEndPuncttrue
\mciteSetBstMidEndSepPunct{\mcitedefaultmidpunct}
{\mcitedefaultendpunct}{\mcitedefaultseppunct}\relax
\EndOfBibitem
\bibitem[Mayer(1983)]{Mayer1983}
I.~Mayer, \emph{Chem. Phys. Lett.}, 1983, \textbf{97}, 270--274\relax
\mciteBstWouldAddEndPuncttrue
\mciteSetBstMidEndSepPunct{\mcitedefaultmidpunct}
{\mcitedefaultendpunct}{\mcitedefaultseppunct}\relax
\EndOfBibitem
\bibitem[Mayer(1986)]{Mayer1986a}
I.~Mayer, \emph{Int. J. Quantum Chem.}, 1986, \textbf{XXIX}, 477--483\relax
\mciteBstWouldAddEndPuncttrue
\mciteSetBstMidEndSepPunct{\mcitedefaultmidpunct}
{\mcitedefaultendpunct}{\mcitedefaultseppunct}\relax
\EndOfBibitem
\bibitem[Alcoba \emph{et~al.}(2008)Alcoba, Bochicchio, Lain, and
  Torre]{Alcoba2008}
D.~R. Alcoba, R.~C. Bochicchio, L.~Lain and A.~Torre, \emph{Phys. Chem. Chem.
  Phys.}, 2008, \textbf{10}, 5144--5146\relax
\mciteBstWouldAddEndPuncttrue
\mciteSetBstMidEndSepPunct{\mcitedefaultmidpunct}
{\mcitedefaultendpunct}{\mcitedefaultseppunct}\relax
\EndOfBibitem
\bibitem[Lobayan \emph{et~al.}(2010)Lobayan, Alcoba, Bochicchio, Torre, and
  Lain]{Lobayan2010}
R.~M. Lobayan, D.~R. Alcoba, R.~C. Bochicchio, A.~Torre and L.~Lain, \emph{J.
  Phys. Chem. A}, 2010, \textbf{114}, 1200--1206\relax
\mciteBstWouldAddEndPuncttrue
\mciteSetBstMidEndSepPunct{\mcitedefaultmidpunct}
{\mcitedefaultendpunct}{\mcitedefaultseppunct}\relax
\EndOfBibitem
\bibitem[Herzberg(1929)]{Herzberg1929}
G.~Herzberg, \emph{Z. Phys.}, 1929, \textbf{57}, 601--630\relax
\mciteBstWouldAddEndPuncttrue
\mciteSetBstMidEndSepPunct{\mcitedefaultmidpunct}
{\mcitedefaultendpunct}{\mcitedefaultseppunct}\relax
\EndOfBibitem
\bibitem[Hall(1987)]{Hall1987}
M.~B. Hall, \emph{Polyhedron}, 1987, \textbf{6}, 679--684\relax
\mciteBstWouldAddEndPuncttrue
\mciteSetBstMidEndSepPunct{\mcitedefaultmidpunct}
{\mcitedefaultendpunct}{\mcitedefaultseppunct}\relax
\EndOfBibitem
\bibitem[Brynda \emph{et~al.}(2006)Brynda, Gagliardi, Widmark, Power, and
  Roos]{Brynda2006}
M.~Brynda, L.~Gagliardi, P.-O. Widmark, P.~P. Power and B.~O. Roos,
  \emph{Angew. Chem. Int. Ed.}, 2006, \textbf{45}, 3804--3807\relax
\mciteBstWouldAddEndPuncttrue
\mciteSetBstMidEndSepPunct{\mcitedefaultmidpunct}
{\mcitedefaultendpunct}{\mcitedefaultseppunct}\relax
\EndOfBibitem
\bibitem[Roos \emph{et~al.}(2007)Roos, Borin, and Gagliardi]{Roos2007}
B.~O. Roos, A.~C. Borin and L.~Gagliardi, \emph{Angew. Chem. Int. Ed.}, 2007,
  \textbf{46}, 1469--1472\relax
\mciteBstWouldAddEndPuncttrue
\mciteSetBstMidEndSepPunct{\mcitedefaultmidpunct}
{\mcitedefaultendpunct}{\mcitedefaultseppunct}\relax
\EndOfBibitem
\bibitem[Nalewajski(2000)]{Nalewajski2000}
R.~F. Nalewajski, \emph{J. Phys. Chem. A}, 2000, \textbf{104},
  11940--11951\relax
\mciteBstWouldAddEndPuncttrue
\mciteSetBstMidEndSepPunct{\mcitedefaultmidpunct}
{\mcitedefaultendpunct}{\mcitedefaultseppunct}\relax
\EndOfBibitem
\bibitem[Nalewajski(2004)]{Nalewajski2004}
R.~F. Nalewajski, \emph{Mol. Phys.}, 2004, \textbf{102}, 531--546\relax
\mciteBstWouldAddEndPuncttrue
\mciteSetBstMidEndSepPunct{\mcitedefaultmidpunct}
{\mcitedefaultendpunct}{\mcitedefaultseppunct}\relax
\EndOfBibitem
\bibitem[Nalewajski(2009)]{Nalewajski2009a}
R.~F. Nalewajski, \emph{J. Math. Chem.}, 2009, \textbf{47}, 692--708\relax
\mciteBstWouldAddEndPuncttrue
\mciteSetBstMidEndSepPunct{\mcitedefaultmidpunct}
{\mcitedefaultendpunct}{\mcitedefaultseppunct}\relax
\EndOfBibitem
\bibitem[Nalewajski and Gurdek(2012)]{Nalewajski2012a}
R.~F. Nalewajski and P.~Gurdek, \emph{Struct. Chem.}, 2012,  1383--1398\relax
\mciteBstWouldAddEndPuncttrue
\mciteSetBstMidEndSepPunct{\mcitedefaultmidpunct}
{\mcitedefaultendpunct}{\mcitedefaultseppunct}\relax
\EndOfBibitem
\bibitem[Barcza \emph{et~al.}(2011)Barcza, Legeza, Marti, and
  Reiher]{orbitalordering}
G.~Barcza, O.~Legeza, K.~H. Marti and M.~Reiher, \emph{Phys. Rev. A}, 2011,
  \textbf{83}, 12508\relax
\mciteBstWouldAddEndPuncttrue
\mciteSetBstMidEndSepPunct{\mcitedefaultmidpunct}
{\mcitedefaultendpunct}{\mcitedefaultseppunct}\relax
\EndOfBibitem
\bibitem[Boguslawski \emph{et~al.}(2012)Boguslawski, Tecmer, Legeza, and
  Reiher]{entanglement_letter}
K.~Boguslawski, P.~Tecmer, O.~Legeza and M.~Reiher, \emph{J. Phys. Chem.
  Lett.}, 2012, \textbf{3}, 3129--3135\relax
\mciteBstWouldAddEndPuncttrue
\mciteSetBstMidEndSepPunct{\mcitedefaultmidpunct}
{\mcitedefaultendpunct}{\mcitedefaultseppunct}\relax
\EndOfBibitem
\bibitem[Boguslawski \emph{et~al.}(2013)Boguslawski, Tecmer, Barcza, Legeza,
  and Reiher]{entanglement_bonding_2013}
K.~Boguslawski, P.~Tecmer, G.~Barcza, {\"O}.~Legeza and M.~Reiher, \emph{J.
  Chem. Theory Comput.}, 2013, \textbf{9}, 2959--2973\relax
\mciteBstWouldAddEndPuncttrue
\mciteSetBstMidEndSepPunct{\mcitedefaultmidpunct}
{\mcitedefaultendpunct}{\mcitedefaultseppunct}\relax
\EndOfBibitem
\bibitem[Tecmer \emph{et~al.}(2014)Tecmer, Boguslawski, Legeza, and
  Reiher]{CUO_DMRG}
P.~Tecmer, K.~Boguslawski, O.~Legeza and M.~Reiher, \emph{Phys. Chem. Chem.
  Phys}, 2014, \textbf{16}, 719--727\relax
\mciteBstWouldAddEndPuncttrue
\mciteSetBstMidEndSepPunct{\mcitedefaultmidpunct}
{\mcitedefaultendpunct}{\mcitedefaultseppunct}\relax
\EndOfBibitem
\bibitem[Rissler \emph{et~al.}(2006)Rissler, Noack, and White]{Rissler_2006}
J.~Rissler, R.~M. Noack and S.~R. White, \emph{Chem. Phys.}, 2006,
  \textbf{323}, 519--531\relax
\mciteBstWouldAddEndPuncttrue
\mciteSetBstMidEndSepPunct{\mcitedefaultmidpunct}
{\mcitedefaultendpunct}{\mcitedefaultseppunct}\relax
\EndOfBibitem
\bibitem[adf()]{adf2012}
ADF2012.01, SCM, Theoretical Chemistry, Vrije Universiteit, Amsterdam, The
  Netherlands, {\tt http://www.scm.com}\relax
\mciteBstWouldAddEndPuncttrue
\mciteSetBstMidEndSepPunct{\mcitedefaultmidpunct}
{\mcitedefaultendpunct}{\mcitedefaultseppunct}\relax
\EndOfBibitem
\bibitem[te~Velde \emph{et~al.}(2001)te~Velde, Bickelhaupt, van Gisbergen,
  Guerra, Baerends, Snijders, and Ziegler]{adf1}
G.~te~Velde, F.~M. Bickelhaupt, S.~J.~A. van Gisbergen, C.~F. Guerra, E.~J.
  Baerends, J.~G. Snijders and T.~Ziegler, \emph{J.~Comput.~Chem.}, 2001,
  \textbf{22}, 931--967\relax
\mciteBstWouldAddEndPuncttrue
\mciteSetBstMidEndSepPunct{\mcitedefaultmidpunct}
{\mcitedefaultendpunct}{\mcitedefaultseppunct}\relax
\EndOfBibitem
\bibitem[Guerra \emph{et~al.}(1998)Guerra, Snijders, te~Velde, and
  Baerends]{adf2}
C.~F. Guerra, J.~G. Snijders, G.~te~Velde and E.~J. Baerends,
  \emph{Theor.~Chem.~Acc.}, 1998, \textbf{99}, 391\relax
\mciteBstWouldAddEndPuncttrue
\mciteSetBstMidEndSepPunct{\mcitedefaultmidpunct}
{\mcitedefaultendpunct}{\mcitedefaultseppunct}\relax
\EndOfBibitem
\bibitem[Perdew(1986)]{Perdew86}
J.~P. Perdew, \emph{Phys. Rev. B}, 1986, \textbf{33}, 8822--8824\relax
\mciteBstWouldAddEndPuncttrue
\mciteSetBstMidEndSepPunct{\mcitedefaultmidpunct}
{\mcitedefaultendpunct}{\mcitedefaultseppunct}\relax
\EndOfBibitem
\bibitem[Becke(1988)]{Becke}
A.~Becke, \emph{Phys.~Rev.~A}, 1988, \textbf{38}, 3098--3100\relax
\mciteBstWouldAddEndPuncttrue
\mciteSetBstMidEndSepPunct{\mcitedefaultmidpunct}
{\mcitedefaultendpunct}{\mcitedefaultseppunct}\relax
\EndOfBibitem
\bibitem[van Lenthe and Baerends(2003)]{adf_b}
E.~van Lenthe and E.~J. Baerends, \emph{J.~Comput.~Chem.}, 2003, \textbf{24},
  1142\relax
\mciteBstWouldAddEndPuncttrue
\mciteSetBstMidEndSepPunct{\mcitedefaultmidpunct}
{\mcitedefaultendpunct}{\mcitedefaultseppunct}\relax
\EndOfBibitem
\bibitem[Roos and Taylor(1980)]{Roos_casscf}
B.~Roos and P.~R. Taylor, \emph{Chem. Phys.}, 1980, \textbf{48}, 157--173\relax
\mciteBstWouldAddEndPuncttrue
\mciteSetBstMidEndSepPunct{\mcitedefaultmidpunct}
{\mcitedefaultendpunct}{\mcitedefaultseppunct}\relax
\EndOfBibitem
\bibitem[Werner and Knowles(1985)]{Werner_1985}
H.-J. Werner and P.~J. Knowles, \emph{J. Chem. Phys.}, 1985, \textbf{82},
  5053--5063\relax
\mciteBstWouldAddEndPuncttrue
\mciteSetBstMidEndSepPunct{\mcitedefaultmidpunct}
{\mcitedefaultendpunct}{\mcitedefaultseppunct}\relax
\EndOfBibitem
\bibitem[Knowles and Werner(1985)]{Knowles_1985}
P.~J. Knowles and H.-J. Werner, \emph{Chem. Phys. Lett.}, 1985, \textbf{115},
  259--267\relax
\mciteBstWouldAddEndPuncttrue
\mciteSetBstMidEndSepPunct{\mcitedefaultmidpunct}
{\mcitedefaultendpunct}{\mcitedefaultseppunct}\relax
\EndOfBibitem
\bibitem[{Werner} \emph{et~al.}(){Werner}, {Knowles}, Lindh, Manby, Sch\"utz,
  Celani, Korona, Mitrushenkov, Rauhut, Adler, and \emph{et al.}]{molpro}
H.-J. {Werner}, P.~J. {Knowles}, R.~Lindh, F.~R. Manby, M.~Sch\"utz, P.~Celani,
  T.~Korona, A.~Mitrushenkov, G.~Rauhut, T.~B. Adler and \emph{et al.},
  \emph{{MOLPRO, Version 2010.1, a Package of \emph{Ab initio} Programs,
  Cardiff University: Cardiff, United Kingdom, and University of Stuttgart:
  Stuttgart, Germany}}\relax
\mciteBstWouldAddEndPuncttrue
\mciteSetBstMidEndSepPunct{\mcitedefaultmidpunct}
{\mcitedefaultendpunct}{\mcitedefaultseppunct}\relax
\EndOfBibitem
\bibitem[Dunning(1989)]{dunning_b}
T.~H. Dunning, \emph{J.~Chem.~Phys.}, 1989, \textbf{90}, 1007\relax
\mciteBstWouldAddEndPuncttrue
\mciteSetBstMidEndSepPunct{\mcitedefaultmidpunct}
{\mcitedefaultendpunct}{\mcitedefaultseppunct}\relax
\EndOfBibitem
\bibitem[Woon and Dunning(1993)]{dunning_bII}
D.~E. Woon and T.~H. Dunning, \emph{J. Chem. Phys}, 1993, \textbf{98},
  1358--1372\relax
\mciteBstWouldAddEndPuncttrue
\mciteSetBstMidEndSepPunct{\mcitedefaultmidpunct}
{\mcitedefaultendpunct}{\mcitedefaultseppunct}\relax
\EndOfBibitem
\bibitem[Legeza()]{dmrg_ors}
O.~Legeza, \emph{\textsc{QC-DMRG-Budapest}, A Program for Quantum Chemical
  {DMRG} Calculations. { \rm Copyright 2000--2013, HAS RISSPO Budapest}}\relax
\mciteBstWouldAddEndPuncttrue
\mciteSetBstMidEndSepPunct{\mcitedefaultmidpunct}
{\mcitedefaultendpunct}{\mcitedefaultseppunct}\relax
\EndOfBibitem
\bibitem[Legeza and S\'{o}lyom(2003)]{legeza_dbss}
O.~Legeza and J.~S\'{o}lyom, \emph{Phys. Rev. B}, 2003, \textbf{68},
  195116\relax
\mciteBstWouldAddEndPuncttrue
\mciteSetBstMidEndSepPunct{\mcitedefaultmidpunct}
{\mcitedefaultendpunct}{\mcitedefaultseppunct}\relax
\EndOfBibitem
\bibitem[Legeza and S\'{o}lyom(2004)]{legeza_dbss3}
O.~Legeza and J.~S\'{o}lyom, \emph{Phys. Rev. B}, 2004, \textbf{70},
  205118\relax
\mciteBstWouldAddEndPuncttrue
\mciteSetBstMidEndSepPunct{\mcitedefaultmidpunct}
{\mcitedefaultendpunct}{\mcitedefaultseppunct}\relax
\EndOfBibitem
\bibitem[Legeza \emph{et~al.}(2003)Legeza, R\"{o}der, and Hess]{legeza_dbss2}
O.~Legeza, J.~R\"{o}der and B.~A. Hess, \emph{Phys. Rev. B}, 2003, \textbf{67},
  125114\relax
\mciteBstWouldAddEndPuncttrue
\mciteSetBstMidEndSepPunct{\mcitedefaultmidpunct}
{\mcitedefaultendpunct}{\mcitedefaultseppunct}\relax
\EndOfBibitem
\bibitem[Altmann and Herzig(1994)]{Altmann}
S.~L. Altmann and P.~Herzig, \emph{Point-Group Theory Tables}, Oxford,
  1994\relax
\mciteBstWouldAddEndPuncttrue
\mciteSetBstMidEndSepPunct{\mcitedefaultmidpunct}
{\mcitedefaultendpunct}{\mcitedefaultseppunct}\relax
\EndOfBibitem
\bibitem[March(1992)]{carbon_book}
J.~March, \emph{Advanced Organic Chemistry: Reactions, Mechanisms, and
  Structure}, Wiley, 1992\relax
\mciteBstWouldAddEndPuncttrue
\mciteSetBstMidEndSepPunct{\mcitedefaultmidpunct}
{\mcitedefaultendpunct}{\mcitedefaultseppunct}\relax
\EndOfBibitem
\bibitem[Abrams and Sherrill(2004)]{Sherill_C2}
M.~L. Abrams and C.~D. Sherrill, \emph{J. Chem. Phys.}, 2004, \textbf{121},
  9211\relax
\mciteBstWouldAddEndPuncttrue
\mciteSetBstMidEndSepPunct{\mcitedefaultmidpunct}
{\mcitedefaultendpunct}{\mcitedefaultseppunct}\relax
\EndOfBibitem
\bibitem[Shaik \emph{et~al.}(2012)Shaik, Danovich, Wu, Su, Rzepa, and
  Hiberty]{Shaik2012}
S.~Shaik, D.~Danovich, W.~Wu, P.~Su, H.~S. Rzepa and P.~C. Hiberty,
  \emph{Nature Chemistry}, 2012, \textbf{4}, 195--200\relax
\mciteBstWouldAddEndPuncttrue
\mciteSetBstMidEndSepPunct{\mcitedefaultmidpunct}
{\mcitedefaultendpunct}{\mcitedefaultseppunct}\relax
\EndOfBibitem
\bibitem[Danovich \emph{et~al.}(2013)Danovich, Shaik, Rzepa, and
  Hoffmann]{Danovich2013}
D.~Danovich, S.~Shaik, H.~S. Rzepa and R.~Hoffmann, \emph{Angew. Chem. Int.
  Ed.}, 2013, \textbf{52}, 5926--5928\relax
\mciteBstWouldAddEndPuncttrue
\mciteSetBstMidEndSepPunct{\mcitedefaultmidpunct}
{\mcitedefaultendpunct}{\mcitedefaultseppunct}\relax
\EndOfBibitem
\bibitem[Frenking and Hermann(2013)]{Frenking2013}
G.~Frenking and M.~Hermann, \emph{Angew. Chem.}, 2013, \textbf{52}, 2--6\relax
\mciteBstWouldAddEndPuncttrue
\mciteSetBstMidEndSepPunct{\mcitedefaultmidpunct}
{\mcitedefaultendpunct}{\mcitedefaultseppunct}\relax
\EndOfBibitem
\bibitem[Matxain \emph{et~al.}(2013)Matxain, Ruip\'{e}rez, Infante, Lopez,
  Ugalde, Merino, and Piris]{Matxain2013}
J.~M. Matxain, F.~Ruip\'{e}rez, I.~Infante, X.~Lopez, J.~M. Ugalde, G.~Merino
  and M.~Piris, \emph{J. Chem. Phys.}, 2013, \textbf{138}, 151102\relax
\mciteBstWouldAddEndPuncttrue
\mciteSetBstMidEndSepPunct{\mcitedefaultmidpunct}
{\mcitedefaultendpunct}{\mcitedefaultseppunct}\relax
\EndOfBibitem
\bibitem[Xu and Dunning(2014)]{Dunning_C2}
L.~T. Xu and T.~H. Dunning, \emph{J. Chem. Theory Comput.}, 2014, \textbf{10},
  195--201\relax
\mciteBstWouldAddEndPuncttrue
\mciteSetBstMidEndSepPunct{\mcitedefaultmidpunct}
{\mcitedefaultendpunct}{\mcitedefaultseppunct}\relax
\EndOfBibitem
\bibitem[Bauschlicher and Langhoff(1987)]{Bauschlicher1987}
C.~W. Bauschlicher and S.~R. Langhoff, \emph{J. Chem. Phys.}, 1987,
  \textbf{87}, 2919\relax
\mciteBstWouldAddEndPuncttrue
\mciteSetBstMidEndSepPunct{\mcitedefaultmidpunct}
{\mcitedefaultendpunct}{\mcitedefaultseppunct}\relax
\EndOfBibitem
\bibitem[Abrams and Sherrill(2004)]{Abrams-C2-2004}
M.~L. Abrams and C.~D. Sherrill, \emph{J. Chem. Phys.}, 2004, \textbf{121},
  9211\relax
\mciteBstWouldAddEndPuncttrue
\mciteSetBstMidEndSepPunct{\mcitedefaultmidpunct}
{\mcitedefaultendpunct}{\mcitedefaultseppunct}\relax
\EndOfBibitem
\bibitem[Sherrill and Piecuch(2005)]{Piecuch-C2-2005}
C.~D. Sherrill and P.~Piecuch, \emph{J. Chem. Phys.}, 2005, \textbf{122},
  124104\relax
\mciteBstWouldAddEndPuncttrue
\mciteSetBstMidEndSepPunct{\mcitedefaultmidpunct}
{\mcitedefaultendpunct}{\mcitedefaultseppunct}\relax
\EndOfBibitem
\bibitem[Evangelista(2011)]{evangelista2011}
F.~A. Evangelista, \emph{J. Chem. Phys.}, 2011, \textbf{134}, 224102\relax
\mciteBstWouldAddEndPuncttrue
\mciteSetBstMidEndSepPunct{\mcitedefaultmidpunct}
{\mcitedefaultendpunct}{\mcitedefaultseppunct}\relax
\EndOfBibitem
\bibitem[Deheng~Shi and Zhu(2011)]{C2-Mol-Phys}
J.~S. Deheng~Shi, Xiaoniu~Zhang and Z.~Zhu, \emph{Mol. Phys.}, 2011,
  \textbf{109}, 1453--1465\relax
\mciteBstWouldAddEndPuncttrue
\mciteSetBstMidEndSepPunct{\mcitedefaultmidpunct}
{\mcitedefaultendpunct}{\mcitedefaultseppunct}\relax
\EndOfBibitem
\bibitem[Wouters \emph{et~al.}(2013)Wouters, Poelmans, Ayers, and
  Neck]{CheMPS2}
S.~Wouters, W.~Poelmans, P.~W. Ayers and D.~V. Neck, \emph{arXiv:1312.2415},
  2013\relax
\mciteBstWouldAddEndPuncttrue
\mciteSetBstMidEndSepPunct{\mcitedefaultmidpunct}
{\mcitedefaultendpunct}{\mcitedefaultseppunct}\relax
\EndOfBibitem
\bibitem[Roskop \emph{et~al.}(2014)Roskop, Kong, Valeev, Gordon, and
  Windus]{Gordon_2014}
L.~B. Roskop, L.~Kong, E.~F. Valeev, M.~S. Gordon and T.~L. Windus, \emph{J.
  Chem. Theory Comput.}, 2014, \textbf{10}, 90--101\relax
\mciteBstWouldAddEndPuncttrue
\mciteSetBstMidEndSepPunct{\mcitedefaultmidpunct}
{\mcitedefaultendpunct}{\mcitedefaultseppunct}\relax
\EndOfBibitem
\bibitem[Boschen \emph{et~al.}(2014)Boschen, Theis, Ruedenberg, and
  Windus]{C2-2014}
J.~S. Boschen, D.~Theis, K.~Ruedenberg and T.~L. Windus, \emph{Theor. Chem.
  Acc.}, 2014, \textbf{133}, 1425\relax
\mciteBstWouldAddEndPuncttrue
\mciteSetBstMidEndSepPunct{\mcitedefaultmidpunct}
{\mcitedefaultendpunct}{\mcitedefaultseppunct}\relax
\EndOfBibitem
\bibitem[Andrews and Wang(2002)]{Lester_Si2H2}
L.~Andrews and X.~Wang, \emph{J. Phys. Chem. A}, 2002, \textbf{106},
  7696--7702\relax
\mciteBstWouldAddEndPuncttrue
\mciteSetBstMidEndSepPunct{\mcitedefaultmidpunct}
{\mcitedefaultendpunct}{\mcitedefaultseppunct}\relax
\EndOfBibitem
\bibitem[Trinquier and Malrieu(1987)]{Si2H4-1}
G.~Trinquier and J.-P. Malrieu, \emph{J. Am. Chem. Soc.}, 1987, \textbf{109},
  5303--5315\relax
\mciteBstWouldAddEndPuncttrue
\mciteSetBstMidEndSepPunct{\mcitedefaultmidpunct}
{\mcitedefaultendpunct}{\mcitedefaultseppunct}\relax
\EndOfBibitem
\bibitem[Malrieu and Trinquier(1989)]{Si2H4-2}
J.-P. Malrieu and G.~Trinquier, \emph{J. Am. Chem. Soc.}, 1989, \textbf{111},
  5916--5921\relax
\mciteBstWouldAddEndPuncttrue
\mciteSetBstMidEndSepPunct{\mcitedefaultmidpunct}
{\mcitedefaultendpunct}{\mcitedefaultseppunct}\relax
\EndOfBibitem
\bibitem[Trinquier(1990)]{Si2H4-3}
G.~Trinquier, \emph{J. Am. Chem. Soc.}, 1990, \textbf{112}, 2130--2131\relax
\mciteBstWouldAddEndPuncttrue
\mciteSetBstMidEndSepPunct{\mcitedefaultmidpunct}
{\mcitedefaultendpunct}{\mcitedefaultseppunct}\relax
\EndOfBibitem
\bibitem[Bogey \emph{et~al.}(1991)Bogey, Bolvin, Demuynck, and
  Destombes]{Si2H2_PRL}
M.~Bogey, H.~Bolvin, C.~Demuynck and J.~L. Destombes, \emph{Phys. Rev. Lett.},
  1991, \textbf{66}, 413--416\relax
\mciteBstWouldAddEndPuncttrue
\mciteSetBstMidEndSepPunct{\mcitedefaultmidpunct}
{\mcitedefaultendpunct}{\mcitedefaultseppunct}\relax
\EndOfBibitem
\bibitem[Lein \emph{et~al.}(2005)Lein, Krapp, and Frenking]{Lein2005}
M.~Lein, A.~Krapp and G.~Frenking, \emph{J. Am. Chem. Soc.}, 2005,
  \textbf{127}, 6290--6299\relax
\mciteBstWouldAddEndPuncttrue
\mciteSetBstMidEndSepPunct{\mcitedefaultmidpunct}
{\mcitedefaultendpunct}{\mcitedefaultseppunct}\relax
\EndOfBibitem
\bibitem[Chesnut(2002)]{Si2H2-good}
D.~B. Chesnut, \emph{Heteroatom. Chem.}, 2002, \textbf{13}, 53--64\relax
\mciteBstWouldAddEndPuncttrue
\mciteSetBstMidEndSepPunct{\mcitedefaultmidpunct}
{\mcitedefaultendpunct}{\mcitedefaultseppunct}\relax
\EndOfBibitem
\bibitem[Adamczyk and Broadbelt(2011)]{Adamczyk2011}
A.~J. Adamczyk and L.~J. Broadbelt, \emph{J. Phys. Chem. A}, 2011,
  \textbf{115}, 2409--2422\relax
\mciteBstWouldAddEndPuncttrue
\mciteSetBstMidEndSepPunct{\mcitedefaultmidpunct}
{\mcitedefaultendpunct}{\mcitedefaultseppunct}\relax
\EndOfBibitem
\bibitem[Snyder and Wasserman(1979)]{Si2H4p1}
L.~C. Snyder and Z.~R. Wasserman, \emph{J. Comput. Chem.}, 1979, \textbf{6},
  5222--5225\relax
\mciteBstWouldAddEndPuncttrue
\mciteSetBstMidEndSepPunct{\mcitedefaultmidpunct}
{\mcitedefaultendpunct}{\mcitedefaultseppunct}\relax
\EndOfBibitem
\bibitem[Poirier and Goddard(1981)]{Si2H4p2}
R.~A. Poirier and J.~D. Goddard, \emph{Chem. Phys. Lett.}, 1981, \textbf{80},
  37--41\relax
\mciteBstWouldAddEndPuncttrue
\mciteSetBstMidEndSepPunct{\mcitedefaultmidpunct}
{\mcitedefaultendpunct}{\mcitedefaultseppunct}\relax
\EndOfBibitem
\bibitem[Lischka and K\"ohler(1982)]{Si2H4p3}
H.~Lischka and H.-J. K\"ohler, \emph{Chem. Phys. Lett.}, 1982, \textbf{85},
  467--471\relax
\mciteBstWouldAddEndPuncttrue
\mciteSetBstMidEndSepPunct{\mcitedefaultmidpunct}
{\mcitedefaultendpunct}{\mcitedefaultseppunct}\relax
\EndOfBibitem
\bibitem[Sax(1985)]{Si2H4p4}
A.~F. Sax, \emph{J. Comput. Chem.}, 1985, \textbf{6}, 469--477\relax
\mciteBstWouldAddEndPuncttrue
\mciteSetBstMidEndSepPunct{\mcitedefaultmidpunct}
{\mcitedefaultendpunct}{\mcitedefaultseppunct}\relax
\EndOfBibitem
\bibitem[Krogh-Jespersen(1985)]{Si2H4p5}
K.~Krogh-Jespersen, \emph{J. Am. Chem. Soc.}, 1985, \textbf{107},
  537--543\relax
\mciteBstWouldAddEndPuncttrue
\mciteSetBstMidEndSepPunct{\mcitedefaultmidpunct}
{\mcitedefaultendpunct}{\mcitedefaultseppunct}\relax
\EndOfBibitem
\bibitem[Frenking and Tonner(2007)]{ABC_bonding}
G.~Frenking and R.~Tonner, \emph{Science}, 2007, \textbf{318}, 746\relax
\mciteBstWouldAddEndPuncttrue
\mciteSetBstMidEndSepPunct{\mcitedefaultmidpunct}
{\mcitedefaultendpunct}{\mcitedefaultseppunct}\relax
\EndOfBibitem
\bibitem[Gier(1961)]{Gier_1961}
T.~E. Gier, \emph{J. Am. Chem. Soc.}, 1961, \textbf{83}, 1769--1770\relax
\mciteBstWouldAddEndPuncttrue
\mciteSetBstMidEndSepPunct{\mcitedefaultmidpunct}
{\mcitedefaultendpunct}{\mcitedefaultseppunct}\relax
\EndOfBibitem
\bibitem[Lucas \emph{et~al.}(2008)Lucas, Michelini, Russo, and
  Sicilia]{Lucas2008}
M.~F. Lucas, M.~C. Michelini, N.~Russo and E.~Sicilia, \emph{J. Chem. Theory.
  Comput.}, 2008, \textbf{4}, 397--403\relax
\mciteBstWouldAddEndPuncttrue
\mciteSetBstMidEndSepPunct{\mcitedefaultmidpunct}
{\mcitedefaultendpunct}{\mcitedefaultseppunct}\relax
\EndOfBibitem
\end{mcitethebibliography}
\bibliographystyle{rsc} %the RSC's .bst file
\providecommand*{\mcitethebibliography}{\thebibliography}
\csname @ifundefined\endcsname{endmcitethebibliography}
{\let\endmcitethebibliography\endthebibliography}{}

\end{document}